\documentclass[12pt]{article}

\usepackage{amsthm,amsmath,amsfonts}
\usepackage{latexsym}
\usepackage[T1]{fontenc}
\usepackage{a4wide}
\usepackage{times}
\usepackage{array}
\usepackage{hhline}
\numberwithin{equation}{section}

\newtheorem{tw}{Theorem}
\newtheorem{lem}{Lemma}
\newtheorem{df}{Definition}

\title{Steady compressible Navier-Stokes flow in a square}
\author{Tomasz Piasecki}

\date{}

\begin{document}

\maketitle

\begin{center}
Mathematical Institute, Polish Academy of Sciences\\
ul. Sniadeckich 8, 00-956 Warszawa, Poland\\
\vskip5mm
e-mail: T.Piasecki@impan.gov.pl\\
\vskip10mm
\textbf{Abstract}
\end{center}

We investigate a steady flow of compressible fluid with inflow boundary condition on the density
and slip boundary conditions on the velocity in a square domain $Q \in \mathbf{R^2}$.
We show existence if a solution $(v,\rho) \in W^2_p(Q) \times W^1_p(Q)$ that is a small
perturbation of a constant flow $(\bar v \equiv [1,0],\bar \rho \equiv 1)$.
We also show that this solution is unique in a class of small perturbations of the constant
flow $(\bar v,\bar \rho)$.
In order show the existence of the solution we adapt the techniques know from the theory
of weak solutions. We apply the method of elliptic regularization and a fixed point argument.

\section{Introduction and main results}

The problems of steady compressible flows described by the Navier-Stokes equations
are usually considered with
the homogeneous Dirichlet boundary conditions on the velocity.
It is worth from the mathematical point of view,
as well as in the eye of applications, to investigate different types of boundary conditions.
A significant feature of the compressible Navier-Stokes system is its mixed character:
the continuity equation is elliptic in the velocity whereas the continuity equation
is hyperbolic in the density.
If we assume that the flow enters the domain, then the hyperbolicity of the continuity equation
makes it necessary
to prescribe the density on the inflow part of the boundary.
A time-dependent compressible flow with inflow boundary condition has been
considered by Valli and Zajaczkowski in \cite{VZ}. The authors showed existence of a global in time solutions under some
smallness assumptions on the data. They also obtained a stability result and existence of a stationary
solution.
Plotnikov and Sokolovski investigated shape optimization problems with inflow boundary condition
in 2D \cite{PS2} and 3D \cite{PS1}, working with weak solutions.
Regular solutions to problems with inflow boundary conditions have been investigated mainly by
Kweon in a joint work with Kellogg \cite{Kw1} and with Song \cite{Kw3}.
The results obtained by these authors require some assumptions
on the geometry of the boundary in the neighbourhood of the points where the inflow and outflow parts of the
boundary meet. In \cite{Kw2} Kweon and Kellogg investigated the case when the inflow and outflow parts of the
boundary are separated, obtaining regular solutions.
What seems to be interesting
is to investigate an inflow condition on the density combined with slip boundary conditions on the velocity,
that allow to describe precisely the action between the fluid and the boundary.
The slip boundary conditions have been investigated
by Mucha \cite{PM1} for incompressible flows, and also by Fujita \cite{Fu} and Mucha and Pokorny \cite{PMMP}
for compressible flows.

Here we investigate a steady flow of a viscous, barotropic, compressible fluid
in a square domain in $\mathbf{R^2}$ satisfying inhomogeneous slip
boundary conditions on the velocity combined with an inflow condition on the density.
We impose that there is no flux across the bottom and the top of the square, so that it can be
considered a finite, two dimensional pipe. From
the analytical point of view our domain prevents the singularity that appears in a general domain
where the inflow and outflow parts of the boundary coincide.

We show existence of a solution that can be considered as a perturbation of a constant
solution $(\bar v \equiv (1,0), \bar \rho \equiv 0)$. Under some smallness assumptions we can show an \emph{a priori}
estimate in a space $W^2_p(Q) \times W^1_p(Q)$ that is crucial in the proof of existence of the solution.
Now let us formulate the problem under consideration more precisely.

The stationary compressible Navier-Stokes system describing the motion of the fluid,
supplied with the slip boundary conditions, reads
\begin{eqnarray}  \label{main_system}
\begin{array}{lcr}
\rho v \cdot \nabla v -\mu \Delta v - (\mu+\nu) \nabla {\rm div}\; v
+\nabla p(\rho) =0 & \mbox{in} & Q,\\
{\rm div}\;(\rho v)=0 & \mbox{in} & Q,\\
%
n\cdot {\bf T}(v,\rho^\gamma)\cdot \tau +f v\cdot \tau=b &
\mbox{on} &\Gamma,\\
n \cdot v=d & \mbox{on} & \Gamma,\\
\rho=\rho_{in} & \mbox{on} & \Gamma_{in},
\end{array}
\end{eqnarray}
where $Q=[0,1]\times[0,1]$ is a square domain in $\mathbf{R^2}$ with the boundary $\Gamma$
and $\Gamma_{in} = \{ x \in \Gamma: \bar v \cdot n(x) < 0 \}$.
We will also denote $\Gamma_{out} = \{ x \in \Gamma: \bar v \cdot n(x) > 0 \}$
and $\Gamma_{0} = \{ x \in \Gamma: \bar v \cdot n(x) = 0 \}$. Next,
$b \in W^{1-1/p}_p(\Gamma), d \in W^{2-1/p}_p(\Gamma)$ and $\rho_{in} \in W^{1-1/p}_p(\Gamma_{in})$
are given functions.
$v=(v^{(1)},v^{(2)})$ is the velocity field of the fluid and $\rho$ is the
density of the fluid. We assume that the pressure is a function of the density
of the form $p(\rho)=\rho^{\gamma}$ for some $\gamma >1$. The outward unit normal and tangent vectors
are denoted respectively by $n$ and $\tau$.
We assume $d=0$ on $\Gamma_0$, what means that there is no flow across these parts
of the boundary.
Moreover,
$$
{\bf T}(v,p) = 2\mu {\bf D}(v) + \nu \, div \,v \,{\bf I} - p {\bf I}
$$
is the stress tensor and
$$
{\bf D}(v) = \frac{1}{2} \{v^i_{x_j}+v^j_{x_i}\}_{i,j=1,2}
$$
is the deformation tensor.
$\mu$ and $\nu$ are viscosity constants satisfying $\mu>0$ and $\nu + 2\mu>0$
and $f>0$ is a friction coefficient. The slip boundary conditions (\ref{main_system})$_{3,4}$
are supplied with the condition (\ref{main_system})$_5$ prescribing the values of the density on the
inflow part of the boundary.
Under the assumptions on $\mu$ and $\nu$ the momentum equation (\ref{main_system})$_{1}$
is elliptic in $u$, whereas the continuity equation (\ref{main_system})$_{2}$ is hyperbolic in $\rho$.

Our method would also work with no modification if we considered
a perturbation of the constant flow $(\bar v,\bar \rho)$ satisfying
(\ref{main_system})$_1$ with a term $\rho F$ on the r.h.s provided that $||F||_{L_p}$ was small enough.

Since $\mathbf{T}(\bar v, \bar \rho^{\gamma})=0$, the constant flow $(\bar v, \bar \rho)$
fulfills equations (\ref{main_system}) with boundary conditions
$f \bar v  \cdot \tau=b - f \tau^{(1)}$ and $n \cdot \bar v = d - f \tau^{(1)}$.

Our main result is
\begin{tw}  \label{main}
Assume that $||b - f \tau^{(1)}||_{W^{1-1/p}_p(\Gamma)}$,$||d - n^{(1)}||_{W^{2-1/p}_p(\Gamma)}$
and $||\rho_{in}-1||_{W^{1-1/p}_p(\Gamma_{in})}$ are small enough and $f$ is large enough.
Then there exists a solution $(v,\rho) \in W^2_p(Q) \times W^1_p(Q)$
to the system (\ref{main_system}) and
\begin{equation} \label{est_main}
||v - \bar v||_{W^2_p} + ||\rho - \bar \rho||_{W^1_p} \leq E,
\end{equation}
where $E$ is a constant depending on the data, i.e. on $d$, $\rho_{in}$, $b$, the constants in the equation
and the domain, that can be arbitrarily small provided that the data is small enough.

Moreover, if $(v_1,\rho_1)$ and $(v_2,\rho_2)$ are two solutions to (\ref{main_system}) satisfying
the estimate (\ref{est_main}) then $(v_1,\rho_1)$ = $(v_2,\rho_2)$.
\end{tw}
There are several difficulties in the proof of Theorem \ref{main} that result, roughly speaking,
from the mixed character of the problem. In a general domain
a singularity appears in the points where the inflow and outflow parts of the boundary
meet and we can not apply the method used in this paper to obtain an \emph{a priori} estimate.
However, there is another difficulty in the analysis of the steady compressible Navier-Stokes system,
independent on the domain. This difficulty lies in the term $u \cdot \nabla w$.
Namely, if we want to apply some fixed point method then
this term makes it impossible to show the compactness of the solution operator.
We overcome this difficulty applying the method of elliptic regularization.
We solve a sequence of approximate elliptic problems and show that this sequence converges to the solution
of (\ref{main_system}). This is a well-known method that has been usually applied to the issue of weak solutions
(\cite{NoS}, \cite{PMMP}), and differs from the approach of Kweon and Kellogg used to derive regular solutions in \cite{Kw1},
\cite{Kw2}.

Let us now outline the strategy of the proof, and thus the structure of the paper.
In section \ref{Prelim} we start with removing inhomogeneity from the boundary conditions
(\ref{main_system})$_{3,4}$.
It leads to the system (\ref{system}), and we can focus on this system instead of (\ref{main_system}).
In the same section we define an $\epsilon$ - elliptic regularization to the system (\ref{system})
and introduce its linearization (\ref{el_linear}).
In section \ref{est_linear} we derive an $\epsilon$ - independent estimate on a solution of the
linearized elliptic system (Theorem \ref{th_w1p}).
Although linear,
the system (\ref{el_linear}) has variable coefficients and thus its solution is not straightforward.
In order to solve (\ref{el_linear}) we apply the Leray-Schauder fixed point theorem in section
\ref{sol_linear}, using a modification of the estimate from Theorem \ref{th_w1p}.
In section \ref{sol_reg} we use the a priori estimate to apply the Schauder fixed point theorem to solve
the approximate elliptic systems. In section \ref{sol_final} we prove our main result, Theorem \ref{main}.
The proof is divided into two steps. First we show that the sequence of approximate
solutions converges to the solution of (\ref{system}) and thus prove the existence of the solution to
(\ref{main_system}) satisfying the estimate (\ref{est_main}). Next we show that this solution is unique
in a class of small perturbations of the constant flow $(\bar v, \bar \rho)$.
We see that the estimate from Theorem \ref{th_el_linear} is in fact used at three stages of the proof,
therefore we show it in a detailed way in section \ref{est_linear}.

\section{Preliminaries} \label{Prelim}
In this section we remove the inhomogeneity
from the boundary conditions (\ref{main_system})$_{4,5}.$
Then we define an $\epsilon$ - elliptic regularization to the system (\ref{main}). We also make some remarks
concerning the notation.
Let us construct $u_0 \in W^2_p(Q)$ and $w_0 \in W^1_p(Q)$ such that
\begin{equation}  \label{extension}
n\cdot u_0|_{\Gamma}= d - n^{(1)} \mbox{ \ \ \ and  \ \ \ }
w_0|_{\Gamma_{in}}= \rho_{in} - 1.
\end{equation}
Due to the assumption of smallness of $d - n^{(1)} |_{\Gamma}$ and $\rho_{in} - 1|_{\Gamma_{in}}$ we can assume
that
\begin{equation} \label{small}
||u_0||_{W^2_p}, \; ||w_0||_{W^2_p} << 1.
\end{equation}
Now we consider
$$
u = v - \bar v - u_0\; \mbox{ \ \ \
and \ \ \ } w = \rho - \bar \rho - w_0.
$$
One can easily verify that $(u,w)$ satisfies the following system:
\begin{eqnarray} \label{system}
\begin{array}{lcr}
\partial_{x_1} u -\mu \Delta  u - (\nu + \mu) \nabla div  u +
\gamma (w + w_0 +1)^{\gamma-1} \nabla  w =  F(u,w) & \mbox{in} & Q,\\
(w + w_0 +1) \, div \, u + \partial_{x_1}w + (u+u_0) \cdot \nabla  w = G(u,w)
& \mbox{in}& Q,\\
n\cdot 2\mu {\bf D}( u)\cdot \tau +f \ u \cdot \tau = B
&\mbox{on} & \Gamma, \\
n\cdot  u = 0 & \mbox{on} & \Gamma,\\
w=0 & \mbox{on} & \Gamma_{in},\\
\end{array}
\end{eqnarray}
where
$$
F(u,w) = -(w + w_0 +1) \, (u_0 \cdot \nabla u + u \cdot \nabla u_0) - w \, (u_0 \cdot \nabla u_0)
$$$$
-(w+w_0+1) \, u \cdot \nabla u - \gamma (w+w_0+1)^{\gamma-1} \nabla w_0
+ \mu \Delta u_0 + (\nu+\mu) \nabla div\, u_0 - (w_0+1) u_0 \cdot \nabla u_0,
$$$$
G(u,w) = -(w + w_0 +1) \, div u_0 - (u + u_0) \cdot \nabla w_0 - \partial_{x_1}w_0
$$
and
$$
B = b - 2 \mu \, n \cdot \mathbf{D}(u_0) \cdot \tau - f \tau^{(1)}.
$$
In order to prove Theorem \ref{main} it is enough to prove the existence of a solution $(u,w)$
to the system (\ref{system}) provided that $||u_0||_{W^2_p}, ||w_0||_{W^1_p}$ and $||B||_{W^{1-1/p}_p(\Gamma)}$
are small enough.
As we already mentioned, the presence of the term $u \cdot \nabla w$ in the continuity
equation makes it impossible to show the compactness of a solution operator if we try to
apply fixed point methods directly to the system (\ref{system}).
We overcome this difficulty applying the method of elliptic regularization.
The method consists of adding an elliptic term $-\epsilon \Delta w$ to the r.h.s of
(\ref{system})$_2$ and introducing an additional Neumann boundary condition.
Since the density is already prescribed on the inflow part of the boundary by (\ref{system})$_5$,
we impose the Neumann condition only on the remaining part of the boundary.
While we are passing to the limit with the density in $W^1_p$ - norm, the Neumann condition will
disappear. Similar approach has been applied to the issue of inviscid limit for the incompressible Euler
system in \cite{GM}.
Consider a following linear system with variable coefficients:
\begin{eqnarray} \label{el_linear}
\begin{array}{lcr}
\partial_{x_1} u_{\epsilon} -\mu \Delta  u_{\epsilon} - (\nu + \mu) \nabla div \, u_{\epsilon} +
\gamma (\bar w +w_0 +1)^{\gamma-1} \nabla  w_{\epsilon} =  F_{\epsilon}(\bar u,\bar w) & \mbox{in} & Q,\\
(\bar w +w_0 +1) \, div \, u_{\epsilon} + \partial_{x_1}w_{\epsilon} + (\bar u+u_0) \cdot \nabla  w_{\epsilon} - \epsilon \Delta w_{\epsilon} = G_{\epsilon}(\bar u,\bar w)
& \mbox{in}& Q,\\
n\cdot 2\mu {\bf D}( u_{\epsilon})\cdot \tau +f \ u_{\epsilon} \cdot \tau = B
&\mbox{on} & \Gamma, \\
n\cdot  u_{\epsilon} = 0 & \mbox{on} & \Gamma,\\
w_{\epsilon}=0 & \mbox{on} & \Gamma_{in},\\
\frac{\partial w_{\epsilon}}{\partial n} = 0 & \mbox{on} & \Gamma \setminus \Gamma_{in}.
\end{array}
\end{eqnarray}
where $(\bar u,\bar w) \in W^2_p(Q) \times W^1_p(Q)$ are given functions
and $F_{\epsilon}(\bar u, \bar w)$ and $G_{\epsilon}(\bar u, \bar w)$ are regularizations to
$F(\bar u, \bar w)$ and $G(\bar u, \bar w)$ obtained by replacing the functions $u_0$ and $w_0$
by their regular approximations $u_0^{\epsilon}$ and $w_0^{\epsilon}$.

Let us define an operator
$T_{\epsilon}: {\cal D} \subset W^2_p(Q) \times W^1_p(Q) \to W^2_p(Q) \times W^1_p(Q):$
\begin{equation} \label{T}
(u_{\epsilon},w_{\epsilon}) = T_{\epsilon}(\bar u, \bar w) \iff (u_{\epsilon},w_{\epsilon}) \quad \textrm{is a solution to (\ref{el_linear})},
\end{equation}
where ${\cal D}$ is a subset of $W^2_p(Q) \times W^1_p(Q)$ that we will define later. Using the
operator $T_{\epsilon}$ we define an $\epsilon$ - elliptic regularization to the system (\ref{system}).

\begin{df}
By an $\epsilon$ - elliptic regularization to the system (\ref{system}) we mean a system
\begin{equation}  \label{el}
(u_{\epsilon},w_{\epsilon}) = T_{\epsilon}(u_{\epsilon},w_{\epsilon}).
\end{equation}
\end{df}

We want to show the existence of a solution to the $\epsilon$ - elliptic regularization
to the system (\ref{system}) applying the Shauder fixed point theorem.
The strategy has been outlined in the introduction.
In section \ref{sol_linear} we show that $T_{\epsilon}$ is well defined, which means that for given $(\bar u,\bar w)$
there exists a unique solution to (\ref{el_linear}) (Theorem \ref{th_el_linear}).
In fact we show
that $T_{\epsilon}$ is well defined for $\epsilon$ small enough, but it suffices since we are interested
in small values of $\epsilon$.

In section \ref{sol_reg} we show that $T_{\epsilon}$ satisfies the assumptions of the Schauder
fixed point theorem and thus we solve the system (\ref{el}) for $\epsilon$ small enough.

As we already said, the key point is to derive an
$\epsilon$ - independent estimate for the system (\ref{el_linear}), which is used
at different stages of the proof. We derive such estimate in the next section. Before we proceed,
we will finish this introductory part with a few remarks concerning notation.

For simplicity we will denote
\begin{equation}  \label{notation}
\begin{array}{c}
\displaystyle
a_0(\bar w) = \frac{\gamma (\bar w+w_0+1)^{\gamma}}{\nu + 2\mu}, \\ [8pt]
a_1(\bar w) = \gamma (\bar w+w_0+1)^{\gamma-1}, \\
a_2(\bar w) = \gamma (\bar w+w_0+1)^{\gamma-2}.
\end{array}
\end{equation}
By $C$ we will denote a constant that depend on the data and thus can be controlled, not necesarily arbitrarily
small. If the constant depend not only on the data, but also on $\epsilon$, we will denote it by $C_{\epsilon}$.
Finally, by $E$ we will denote a constant dependent on the data that can be arbitrarily small provided that
the data is small enough.

Since we will usually use the spaces of functions defined
on $Q$, we will omit $Q$ in the notation of a space, for example we will denote the space $L_2(Q)$ by $L_2$.
The spaces of functions defined on the boundary will be denoted by $L_2(\Gamma)$ etc.

We do not distinguish between the spaces of vector-valued and scalar-valued functions, for example
we will write $u \in W^2_p$ instead of $u \in (W^2_p)^2$.

\section{A priori estimate for the linearized elliptic system} \label{est_linear}
In this section we show an
$\epsilon$ - independent estimate on $||u_{\epsilon}||_{W^2_p} + ||w_{\epsilon}||_{W^1_p}$,
where $(u_{\epsilon},w_{\epsilon})$ is a solution to (\ref{el_linear}).
The first step is an estimate in $H^1 \times L_2$. Next we eliminate
the term $div \,u$ from the second equation applying the Helmholtz decomposition and the properties
of the slip boundary conditions. Then we derive the higher estimate using interpolation.
\subsection{Estimate in $H^1 \times L_2$}  \label{enefirst}
In order to prove a priori estimates on $H^1$ - norm of the
velocity and $L^2$ - norm of the density for the system (\ref{el_linear})
let us define a space
\begin{equation}  \label{v}
V = \{ v \in H^1(Q;\mathbf{R^2}): v \cdot n|_{\Gamma} = 0\}.
\end{equation}
The estimate is stated in the following lemma.
\begin{lem} \label{lem_ene1}
Assume that $\epsilon$, $||\bar u||_{W^2_p}$ and $||\bar w||_{W^1_p}$ are small enough and $f$ is large enough.
Then for sufficiently smooth solutions to system (\ref{el_linear}) the
following estimate is valid
\begin{equation} 	\label{ene1}
||u||_{W^1_2}+||w||_{L_2} \leq C \big[ ||F(\bar u, \bar w)||_{V^*} + ||G(\bar u, \bar w)||_{L_2} + ||B||_{L^2(\Gamma)} + E ||w||_{W^1_p} \big].
\end{equation}
where $V^*$ is the dual space of $V$.
\end{lem}

Before we start the proof, we shall make a remark concerning the term $||w||_{W^1_p}$, that
is rather unexpected in an energy estimate. Its presence is due to the functions
$a_1(\bar w)$ and $(\bar w + w_0 +1)$ on the r.h.s. of (\ref{el_linear}).
However, this term does not cause any problems
when we apply (\ref{ene1}) to interpolate in the proof of Theorem \ref{th_w1p},
since it is multiplied by a small constant.

\textbf{Proof.}
The proof is divided into three steps. First we multiply (\ref{el_linear})$_1$
by $u$ and integrate over $Q$. We obtain an estimate on $||u||_{H^1}$ in terms of the data and $||w||_{L_2}$.
Then we apply the second equation to estimate $||w||_{L_2}$ and finally combine these estimates to obtain
(\ref{ene1}).

\textbf{Step 1.} We multiply (\ref{el_linear})$_1$ by $u$ and integrate over $Q$.
Using the boundary conditions (\ref{el_linear})$_{3,4}$ we get
\begin{equation} \label{a5}
\begin{array}{c}
\int_Q 2\mu{\bf D}^2(u) + \nu div^2u \,dx + \int_{\Gamma} (f + \frac{n^{(1)}}{2}) |u|^2 \,d\sigma
+ \int_Q [ a_1(\bar w) ] \nabla w u dx = \\
+\int_{Q} F u dx + \int_{\Gamma} B(u \cdot \tau) \, d\sigma.
\end{array}
\end{equation}
The boundary term on the l.h.s will be positive provided that $f$ is large enough.
Next we integrate by parts the last term of the l.h.s of (\ref{a5}).
Using (\ref{el_linear})$_2$ we obtain:
$$
\int_{Q} [ a_1(\bar w) ] \nabla w u dx= - \int_{Q} [ a_1(\bar w) ] \, div\;u w dx - \int_Q u \, w \, \nabla [ a_1(\bar w) ] \,dx  =
$$$$
= \int_{\Gamma} \frac{[ a_2(\bar w) ]}{2} w^2 n^{(1)} \,d\sigma -\frac{1}{2} \int_Q w^2 \, [ \partial_{x_1} a_2(\bar w) + (\bar u+u_0) \nabla a_2(\bar w) ] \,dx  - \frac{1}{2} \int_{Q} [ a_2(\bar w) ] div\, (\bar u + u_0) \,w^2 \,dx
$$$$
- \int_{Q} [ a_2(\bar w) ] G(\bar u,\bar w) w \,dx
- \epsilon \int_{Q} [ a_2(\bar w) ] w \Delta w \,dx - \int_Q u \, w \, \nabla [ a_1(\bar w) ] \,dx.
$$
Since $n^{(1)}|_{\Gamma_{out}}\equiv 1$,
using (\ref{a5}) and the Korn inequality ((\ref{Korn}), Appendix) we get:
\begin{equation}  \label{a7}
\begin{array}{c}
C_Q || u ||_{W_2^1}^2 + \int_{\Gamma_{out}} [ a_2(\bar w) ] \, w^2 \, d\sigma  \leq \\
\leq \underbrace{ \int_{Q} a_2(\bar w) \, div \, (\bar u +u_0) w^2 \, dx}_{I_1} + \underbrace{ \int_{Q} [ a_2(\bar w) ] \, Gw \,dx + \int_{Q} Fu \,dx + \int_{\Gamma} B (u \cdot \tau) \, d\sigma}_{I_2} \\
+ \underbrace{ \epsilon \int_Q [ a_2(\bar w) ] w \Delta w \,dx}_{I_3} + \underbrace{ \int_Q u \, w \, \nabla [ a_1(\bar w) ] \,dx}_{I_4} + \underbrace{ \int_Q w^2 \, [ \partial_{x_1} a_2(\bar w) + (\bar u+u_0) \nabla a_2(\bar w) ] \,dx}_{I_5}.
\end{array}
\end{equation}

Obviously we have $I_1 \leq E \, ||w||_{L_2}$.
Now we have to deal with the term with $\Delta w$. Due to the boundary conditions (\ref{el_linear})$_{5,6}$ we have
\begin{equation}
I_3 = \epsilon \int_Q [ a_2(\bar w) ] w \Delta w \,dx = - \epsilon \int_{Q} [ a_2(\bar w) ] \, |\nabla w|^2 \,dx - \epsilon \int_Q w \, \nabla [ a_2(\bar w) ] \, \nabla w \,dx.
\end{equation}
Using H\"older inequality we get
$$
| \int_Q w \, \nabla [ a_2(\bar w) ] \, \nabla w \,dx | \leq ||\nabla [ a_2(\bar w) ]||_{L_p} ||w \, \nabla w||_{L_{p^*}} \leq
$$$$
\leq ||\nabla [ a_2(\bar w) ]||_{L_p} ||\nabla w||_{L_2} \, ||w||_{L_q} \leq C \, ||\nabla w||_{L_2}^2,
$$
where $q = \frac{2p}{p-2} < +\infty$ and $p^*=\frac{p}{p-1}$. Thus the term with $\epsilon$ on the r.h.s of (\ref{a7}) will be negative provided
that $||\bar w||_{W^1_p}$ will be small enough.
Next,
$$
I_4 \leq |\int_Q u \, w \, \nabla [ a_1(\bar w) ] \,dx| \leq C ||\nabla [ a_1(\bar w) ]||_{L_p} \, ||u||_{W^1_2} \, ||w||_{L_2}
\leq E \big( ||u||_{W^1_2}^2 + ||w||_{L_2}^2 \big).
$$
The last term of the r.h.s. is the most inconvenient and it must be estimated by $W^1_p$ - norm of $w$,
and this is the reason why this term appears in (\ref{ene1}).
Fortunately it is multiplied by a small constant what will turn out very important in the proof of Theorem \ref{th_w1p}.
We have
$$
I_5 \leq C \, ||a_2(\bar w)||_{W^1_p} \, ||w||_{W^1_p}^2 \leq E \, ||w||_{W^1_p}^2.
$$
Provided that the data is small enough, using the trace theorem to estimate the boundary term and the H\"older inequality we get
\begin{equation} \label{u1}
\begin{array}{c}
||u||^2_{W^1_2} + C \, \int_{\Gamma_{out}} w^2 \,d\sigma \leq \\
\leq C \, \big[ ||F(\bar u, \bar w)||_{V^*} + ||G(\bar u, \bar w)||_{L_2} + ||B||_{L_2(\Gamma)} \big] (||u||_{W^1_2} + ||w||_{L_2})
+ E\,||w||_{W^1_p}^2.
\end{array}
\end{equation}
%
%

\textbf{Step 2.} In order to derive (\ref{ene1}) from (\ref{u1}) we need to find a bound on $||w||_{L_2}$.
From $(\ref{el_linear})_2$ we have
$$
\partial_{x_1}w= G-(\bar u+u_0)\cdot \nabla w - (\bar w + w_0 +1) \, div \;u + \epsilon \Delta w,
$$
thus
$$
w^2(x_1,x_2) = w^2(0,x_2) + \int_{0}^{x_1}2 w \, w_s(s,x_2) \,ds =
$$$$
\underbrace{\int_{0}^{x_1} 2 w [G - (\bar w + w_0 +1) \, div \,u] \,ds}_{S_1}
\underbrace{-\int_{0}^{x_1}2 w_2 (\bar u+u_0) \cdot \nabla w_2 \,ds + 2\epsilon \int_{0}^{x_1} w_2 \Delta w_2 \,dx}_{S_2}.
$$
$S_1$ can be estimated directly:
\begin{equation}    \label{w1}
\int_Q S_1 \leq (||G||_{L_2} + C ||u||_{H_1}) ||w||_{L_2}.
\end{equation}
It is a little more complicated to estimate $S_2$. We have
$$
S_2 = -\int_{0}^{x_1} (\bar u+u_0)^{(1)} \partial_s w^2(s,x_2) \,ds -
\int_{0}^{x_1} (\bar u+u_0)^{(2)} \partial_{x_2} w^2 (s,x_2) \,ds + 2\epsilon \int_{0}^{x_1} w \Delta w.
$$
Now we integrate both components by parts. In the second component we use the fact
that the integration interval does not depend on $x_2$. We get
%
%
%
$$
S_2 = - (\bar u+u_0)^{(1)}w^2(x_1,x_2) + \int_{0}^{x_1}(\bar u+u_0)^{(1)}_{x_1} w^2(s,x_2) \,ds
$$$$
- \frac{\partial}{\partial x_2} \int_{0}^{x_1} (\bar u+u_0)^{(2)}w^2(s,x_2) \,ds
+ \int_{0}^{x_1} (\bar u+u_0)^{(2)}_{x_2} w^2(s,x_2) \,ds + 2\epsilon \int_{0}^{x_1} w \Delta w =
$$$$
= - (\bar u+u_0)^{(1)}w^2(x_1,x_2) + \int_{0}^{x_1} w^2 \, div \;(\bar u+u_0) (s,x_2) \,ds -
\frac{\partial}{\partial x_2} \int_{0}^{x_1} (\bar u+u_0)^{(2)}w^2(s,x_2) \,ds +
$$$$
2\epsilon \int_{0}^{x_1} w \Delta w =: S_2^1 + S_2^2 + S_2^3 + S_2^4.
$$
The integrals of
$S_2^1$ and $S_2^2$ can be estimated in a direct way:
\begin{equation} \label{i1i2}
\int_{Q}|S_2^1|, \int_{Q}|S_2^2| \leq E \, ||w||_{L^2}^2.
\end{equation}
Next,
$$
\int_{Q} S_2^3 = \int_{Q} \frac{\partial}{\partial x_2} \big[ \int_{0}^{x_1} u^{(2)}w^2(s,x_2)\,ds \big] \,dx
= \int_{\Gamma} n^{(2)} \big[ \int_{0}^{x_1} (\bar u+u_0)^{(2)}w^2(s,x_2)\,ds \big] \,d\sigma.
$$
Now we remind that $w=0$ on $\Gamma_{in}$. Moreover, the boundary conditions yields
$(\bar u+u_0)^{(2)} =0$ on $\Gamma_0$.
Finally, on $\Gamma_{out}$ we have $n^{(2)}=0$. Thus
\begin{equation}  \label{i3}
\int_{Q} S_2^3 = 0.
\end{equation}
Finally,
$$
\int_Q S_2^4 \,dx = \int_{0}^{1} \big[ \int_{0}^{1} \int_{0}^{x_1} w \Delta w (s,x_2) \,ds \,dx_2 \big] \,dx_1 =
\int_0^1 \big[ \int_{P_{x_1}} w \Delta w(x) \,dx \big] \,dx_1,
$$
where $P_{x_1} := [0,x_1] \times [0,1]$. We have
$$
\int_{P_{x_1}} w \Delta w \,dx = -\int_{P_{x_1}} |\nabla w|^2 \,dx + \int_{\partial P_{x_1}} w \, \nabla w \cdot n \,d\sigma
\overset{(\ref{el_linear})_{5,6}}{\leq} \int_0^1 w w_{x_1} (x_1,x_2) \,dx_2,
$$
thus
\begin{equation} \label{i4}
\int_Q S_2^4 \,dx \leq 2\epsilon \int_0^1 \int_0^1 w w_{x_1}(x_1,x_2) \,dx_2 \,dx_1 =
\epsilon \int_Q \partial_{x_1} w^2 \,dx = \epsilon \int_{\Gamma_{out}} w^2 n^{(1)} \,d\sigma.
\end{equation}
Combining (\ref{i1i2}), (\ref{i3}) and (\ref{i4}) we get
$$
\int_{Q} S_2 = \int_Q S_2^1 + S_2^2 + S_2^3 + S_2^4 \leq E \, ||w||_{L^2}^2
+ \epsilon \int_{\Gamma_{out}} w^2 \,d\sigma.
$$
Combining this estimate with (\ref{w1}) 
we get:
$$
||w||_{L^2}^2 \leq C \big( ||G(\bar u,\bar w)||_{L_2} + ||u||_{W^1_2} \big)^2
+ E \, ||w||_{L^2} + \epsilon \int_{\Gamma_{out}} w^2 \,d\sigma,
$$
and thus
\begin{equation}  \label{wl2}
||w||_{L^2}^2 \leq C \big( ||G(\bar u,\bar w)||_{L_2} + ||u||_{W^1_2} \big)^2 + C \, \epsilon \int_{\Gamma_{out}} w^2 \,d\sigma.
\end{equation}
\textbf{Step 3.} Substituting (\ref{wl2}) to (\ref{u1}) we get:
\begin{equation} \label{u2}
||u||_{W^1_2}^2 + \int_{\Gamma_{out}} w^2 \,d\sigma \leq C \, D (||u||_{W^1_2} + ||w||_{L_2}) + C \, D^2 + E \, ||w||_{W^1_p}^2,
\end{equation}
where $D=||F(\bar u , \bar w)||_{V^*} + ||G(\bar u , \bar w)||_{L^2} + ||B||_{L^2(\Gamma)}$. Combining this inequality with (\ref{wl2}) we get
$$
(||u||_{W^1_2} + ||w||_{L_2})^2 + (C-\epsilon) \int_{\Gamma_{out}} w^2 \,d\sigma \leq
C \, D \, (||u||_{W^1_2} + ||w||_{L_2}) + D^2 + E \, ||w||_{W^1_p}^2 ,
$$
thus for $\epsilon$ small enough we obtain (\ref{ene1}). $\square$
\subsection{Estimate for $||u||_{W^2_p}+||w||_{W^1_p}$}
The following theorem gives an $\epsilon$ - independent estimate on $||u_{\epsilon}||_{W^2_p} + ||w_{\epsilon}||_{W^1_p}$
where $(u_{\epsilon},w_{\epsilon})$ is a solution to (\ref{el_linear}).
\begin{tw} \label{th_w1p}
Suppose that $(u_{\epsilon},w_{\epsilon})$ is a solution to (\ref{el_linear}).
Then the following estimate is valid
provided that the data, $||\bar u||_{W^2_p}$ and $||\bar w||_{W^1_p}$ are small enough
and $f$ is large enough.
\begin{equation} \label{est_el_linear}
||u_{\epsilon}||_{W^2_p} + ||w_{\epsilon}||_{W^1_p} \leq C \, \big[ ||F_{\epsilon}(\bar u,\bar w)||_{L_p} + ||G_{\epsilon}(\bar u,\bar w)||_{W^1_p} + ||B||_{W^{1-1/p}_p(\Gamma)} \big],
\end{equation}
where the constant $C$ depends on the data but does not depend on $\epsilon$.
\end{tw}
The proof will be divided into three lemmas.
In the first lemma we eliminate the term $div \, u$ from (\ref{el_linear})$_2$.
\begin{lem}
Let us define
\begin{equation}
\bar H := -(\nu+2\mu) \, div \,u_{\epsilon} + [ a_1(\bar w) ] w_{\epsilon}. \label{div1}
\end{equation}
where $(u_{\epsilon}, w_{\epsilon})$ is a solution to (\ref{el_linear})
and $a_1(\bar w)$ is defined in (\ref{notation}). Then
\begin{equation}  \label{nablahlp}
||\nabla \bar H||_{L_p}
\leq C \Big [||F_{\epsilon}(\bar u,\bar w)||_{L_p} + ||B||_{W^{1-1/p}_p(\Gamma)} + ||u||_{W^{1-1/p}_p(\Gamma)} \Big] + E \, ||w||_{W^1_p}
\end{equation}
and $w_{\epsilon}$ satisfies the following equation
\begin{equation}  \label{w}
[ a_0(\bar w) ] w + w_{x_{1}} + (\bar u+u_0) \cdot \nabla w - \epsilon \Delta w = \tilde H,
\end{equation}
where
\begin{equation}
\tilde H = \frac{\bar H \, (\bar w+w_0+1) }{\nu + 2\mu} +G(\bar u,\bar w).  \label{tildeh}
\end{equation}
\end{lem}
\textbf{Proof.}
Let us rewrite (\ref{el_linear})$_1$ as
$$
\partial_{x_1} u_{\epsilon} -\mu \Delta  u_{\epsilon} - (\nu + \mu) \nabla div \, u_{\epsilon} +
\gamma \nabla w_{\epsilon}  =  F_{\epsilon}(\bar u,\bar w) - [a_1(\bar w) - \gamma] \nabla w_{\epsilon}.
$$
Taking the two dimensional vorticity of (\ref{el_linear})$_1$ we get
\begin{equation} \label{alpha}
\begin{array}{lcr}
\partial_{x_1} \alpha_{\epsilon}- \mu \Delta \alpha_{\epsilon} = rot\, [ F_{\epsilon}(\bar u,\bar v) - (a_1(\bar w)-\gamma) \nabla w_{\epsilon}] & \mbox{in} & Q, \\
\alpha_{\epsilon}=- \frac{f}{\mu} (u_{\epsilon} \cdot \tau) + \frac{B}{\mu} & \mbox{on} & \Gamma,
\end{array}
\end{equation}
where $\alpha_{\epsilon}=rot\; u_{\epsilon} = u^{(2)}_{\epsilon , x_1}-u^{(1)}_{\epsilon,x_2}$.
The boundary condition (\ref{alpha})$_2$ has been shown in \cite{PM1} in a more general case;
a simplification of this proof yields (\ref{alpha})$_2$.
%
Since our domain is a square, we can use the symmetry to deal with corner singularites
and apply the standard $L^p$ theory of elliptic equations (\cite{Gi}) to obtain the estimate
\begin{equation}  \label{rotuw1p_1}
||\alpha_{\epsilon}||_{W_p^1} \leq C \big[ ||F_{\epsilon}(\bar u,\bar w)||_{L_p(Q)} + ||(a_1(\bar w)-\gamma) \nabla w_{\epsilon}||_{L_p}
+||- \frac{f}{\mu} (u_{\epsilon} \cdot \tau) + \frac{B}{\mu}||_{W^{1-1/p}_p(\Gamma)} \big] .
\end{equation}
From the definition of $a_1(\bar w)$ (\ref{notation}) we see that $||(a_1(\bar w)-\gamma)||_{L_\infty}$
can be arbitrarily small provided that $||\bar w||_{W^1_p}$ is small enough. Moreover, from the boundary condition
(\ref{el_linear})$_{4}$ we have $u_{\epsilon} = \tau (u_{\epsilon} \cdot \tau)$ on $\Gamma$, thus
(\ref{rotuw1p_1}) can be rewritten as
\begin{equation}  \label{rotuw1p}
||\alpha_{\epsilon}||_{W_p^1} \leq C \big[ ||F_{\epsilon}(\bar u,\bar w)||_{L_p(Q)} + ||B||_{W^{1-1/p}_p(\Gamma)}
+ ||u_{\epsilon}||_{W^{1-1/p}_p(\Gamma)} + E \, ||w_{\epsilon}||_{W^1_p} \big].
\end{equation}
Now we apply the Helmholtz decomposition in of $u_{\epsilon}$ (see Appendix, (\ref{Helm})):
\begin{equation}
u_{\epsilon} = \nabla \phi + \nabla^{\perp} A.
\end{equation}
For simplicity we omit the index $\epsilon$ in the notation of $\phi$ and $A$.
We have $n \cdot \nabla^{\perp} A = \tau \cdot \nabla A = \frac{\partial}{\partial \tau}A$,
thus the condition $n \cdot \nabla^{\perp} A|_{\Gamma} =0$ yields
$A|_{\Gamma} = \textrm{const}$. Moreover,
$$
rot \,u = rot(\nabla \phi + \nabla^{\perp} A) =
rot \, \nabla^{\perp} A = \Delta A.
$$
We see that $A$ is a solution to the following boundary value problem:
\begin{eqnarray*}
\left\{ \begin{array}{c}
\Delta A = \alpha_{\epsilon} \in W^1_p(Q), \\
A|_{\Gamma} = \textrm{const}.
\end{array} \right.
\end{eqnarray*}
Applying again the elliptic theory we get
\begin{equation} \label{Aw3p}
||A||_{W^3_p(Q)} \leq ||\alpha||_{W^1_p(Q)} \leq
C \left\{ ||F_{\epsilon}(\bar u,\bar w)||_{L_p(Q)} +
||B||_{W^{1-1/p}_p(\Gamma)} + ||u_{\epsilon}||_{W^{1-1/p}_p(\Gamma)}
\right\} .
\end{equation}
Substituting the Helmholtz decomposition (\ref{Helm}) to (\ref{el_linear})$_1$ we get
$$
\partial_{x_{1}}(\nabla \phi + \nabla^{\perp} A)
-\mu \Delta (\nabla \phi + \nabla^{\perp} A)
-(\nu+\mu) \nabla div (\nabla \phi + \nabla^{\perp} A) + [ a_1(\bar w) ]  \, \nabla w = F_{\epsilon}(\bar u,\bar w),
$$
but $div \nabla \phi = \Delta \phi$ and thus
\begin{equation} \label{barF}
\begin{array}{c}
-(\nu + 2\mu) \nabla \Delta \phi + \nabla ( [ a_1(\bar w) ] w ) = \\
=F(\bar u,\bar w) + \mu \Delta \nabla^{\perp} A + (\mu+\nu) \nabla div \nabla^{\perp} A - \partial_{x_{1}}\nabla^{\perp} A + \partial_{x_{1}} \nabla \phi + w \nabla [ a_1(\bar w) ]
=: \bar F,
\end{array}
\end{equation}
what can be rewritten as:
$$
\nabla \left( -(\nu+2\mu)\Delta \phi + \gamma a_1(\bar w) w \right) = \bar F.
$$
We have $\Delta \phi = div \,u$, thus $\bar F = \nabla \bar H$ where $\bar H$ is defined in (\ref{div1}).
From (\ref{barF}) we have
\begin{displaymath}
\begin{array}{c}
||\bar F||_{L_p} \leq C [||F_{\epsilon}(\bar u,\bar w)||_{L^p} + ||A||_{W^3_p} + ||\nabla ^2 \phi||_{L_p}] + ||\nabla [ a_0(\bar w) ]||_{L_p} ||w_{\epsilon}||_{\infty} \leq
\\
\leq C [||F_{\epsilon}(\bar u,\bar w)||_{L^p} + ||A||_{W^3_p} + ||\phi||_{W^2_p}] + E \, ||w_{\epsilon}||_{W^1_p}
\end{array}
\end{displaymath}
and from (\ref{Aw3p}) and (\ref{est_helm}) we get (\ref{nablahlp}). The proof is thus completed. $\square$

In the next lemma we will use the equation (\ref{w}) to estimate $||w||_{W^1_p}$ in the terms of functions
$\bar H$ and $G(\bar u, \bar w)$.

\begin{lem}
Under the assumptions of Theorem \ref{th_w1p} the following estimate is valid:
\begin{equation} \label{w-par-w1p}
||w_{\epsilon}||_{W^1_p} + ||w_{\epsilon,x_1}||_{L_p(\Gamma_{in})}
\leq C \, \big[ ||H||_{W^1_p} + ||H|_{\Gamma_{in}}||_{L_p(\Gamma_{in})} \big] ,
\end{equation}
where
\begin{equation}  \label{h}
H = \frac{\bar H}{\nu + 2\mu} +G(\bar u,\bar w).
\end{equation}
\end{lem}
\bigskip
\textbf{Proof.}
Throughout the proof we will omit the index $\epsilon$ denoting $w_{\epsilon}$ by $w$.
The proof will be divided into four steps. First we estimate $||w||_{L_p}$, then $||w_{x_1}||_{L_p}$
and $||w_{x_2}||_{L_p}$ and finally combine these estimates.

\textbf{Step 1.} Multiplying (\ref{w}) by $|w|^{p-2} \, w$ and integrating over $Q$ we get:
\begin{equation} \label{w_multiplied}
\underbrace{\int_Q a_0(\bar w) \, |w|^p}_{I_1^1} + \underbrace{\int_Q |w|^{p-2}w \,w_{x_{1}}}_{I_1^2} + \underbrace{\int_Q (\bar u+u_0) \cdot \nabla w |w|^{p-2}w}_{I_1^3} - \underbrace{\epsilon \int_Q \Delta w |w|^{p-2}w}_{I_1^4}
= \underbrace{\int_Q \tilde H |w|^{p-2}w}_{I_1^5}.
\end{equation}
We have
$$
I_1^3 = \frac{1}{p} \int_{Q} (u+u_0) \cdot \nabla |w|^p \,dx =
-\frac{1}{p} \int_{Q} |w|^p div \,(\bar u+u_0) \,dx + \frac{1}{p} \int_{\Gamma_{out}} u_0^{(1)} |w|^p d\sigma.
$$
Next,
$$
I_1^2 = \frac{1}{p} \int_Q \partial_{x_1} |w|^p
= \frac{1}{p} \int_{\Gamma} |w|^p \, n^{(1)} = \frac{1}{p} \int_{\Gamma_{out}} |w|^p.
$$
Combining the last two equations we get
$$
-(I_1^2 + I_1^3) \leq E \, ||w||_{L_p}^p - C \, \int_{\Gamma_{out}} (1+u_0^{(1)}) |w|^p \,d\sigma.
$$
The boundary term is positive due to the assumption of smallness of $u_0$.
The term with $\Delta w$:
$$
I_1^4 =
\epsilon \int_Q \nabla w \cdot \nabla (|w|^{p-2}w) -
\epsilon \int_{\Gamma} |w|^{p-2}w \frac{\partial w}{ \partial n}.
$$
The boundary term vanishes due to the conditions (\ref{el_linear})$_{5,6}$
and the first term of the r.h.s is equal to
$$
C \, \int_Q ( |\nabla w|^{p-2} \, w^2 + |w|^{p-2} |\nabla w|^2 ) dx \geq 0.
$$
The r.h.s of (\ref{w_multiplied}) can be estimated directly:
$$
I_1^5 = \int_{Q} \tilde H |w|^{p-2} w \,dx \leq
||\tilde H||_{L_p} \int_{Q} \big( |w|^{(p-1)p^*} \big)^{1/p^*} =
||\tilde H||_{L_p} ||w||_{L_p}^{p-1}.
$$
The smallness of $\bar w$ and $w_0$ in $W^1_p$ implies that $a_0(\bar w) \geq C >0$,
thus combining the above estimates we get
$C \, ||w||_{L_p}^p \leq ||\tilde H||_{L_p} ||w||_{L_p}^{p-1} + E \, ||w||_{L_p}^p$,
thus
\begin{equation} \label{wlp}
||w||_{L_p} \leq C \, ||\tilde H||_{L_p}.
\end{equation}

\textbf{Step 2.} In order to estimate $w_{x_2}$ we differentiate (\ref{w}) with respect to $x_2$,
multiply it by $|w_{x_2}|^{p-2} w_{x_2}$ and integrate over $Q$. We get
$$
\underbrace{\int_{Q} [ a_0(\bar w) ] |w_{x_2}|^p}_{I_2^1} + \underbrace{\int_Q [ a_0(\bar w) ]_{x_2} w |w_{x_2}|^p w_{x_2}}_{I_2^2} + \underbrace{\int_Q |w_{x_2}|^{p-2} w_{x_2} w_{x_2 x_1}}_{I_2^3}
$$$$
+ \underbrace{\int_Q ((\bar u+u_0)_{x_2} \cdot \nabla w) \, |w_{x_2}|^{p-2} w_{x_2}}_{I_2^4} +
\underbrace{\int_Q ((\bar u+u_0) \cdot \nabla w_{x_2}) \, |w_{x_2}|^{p-2} w_{x_2}}_{I_2^5}
$$$$
-\underbrace{ \epsilon \int_Q \Delta w_{x_2} |w_{x_2}|^{p-2} \, w_{x_2} }_{I_2^6}
= \underbrace{\int_{Q} \tilde H_{x_2} |w_{x_2}|^{p-2} w_{x_2}}_{I_2^7}.
$$
We have
$$
I_2^3 = \frac{1}{p} \int_{Q} \partial_{x_1} |w_{x_2}|^p \,dx
= - \frac{1}{p} \int_{\Gamma_{in}} |w_{x_2}|^p \,d\sigma + \frac{1}{p} \int_{\Gamma_{out}} |w_{x_2}|^p \,d\sigma,
$$
but the condition $w=0$ on $\Gamma_{in}$ implies $w_{x_2}=0$ on $\Gamma_{in}$, thus
\begin{equation}	\label{wx2_1}
I_2^3 = \frac{1}{p} \int_{\Gamma_{out}} |w_{x_2}|^p \,d\sigma.
\end{equation}
Obviously we have $I_2^4 \leq E \, ||\nabla w||_{L_p}^p$.
Next,
$$
I_2^5 = - \frac{1}{p} \int_{Q} div(\bar u+u_0) |w_{x_2}|^p + \frac{1}{p} \int_{\Gamma_{out}} u_0^{(1)} n^{(1)} |w_{x_2}|^p.
$$
Combining this equation with (\ref{wx2_1}) we get
\begin{displaymath}
I_2^3 + I_2^5 =
-1/p \int_{Q} div(\bar u+u_0) |w_{x_2}|^p + \frac{1}{p} \int_{\Gamma_{out}} (1+u_0^{(1)}) |w_{x_2}|^p \,d\sigma,
\end{displaymath}
The boundary term is nonnegative due to the smallness of $u_0$.
\\
The last part of the l.h.s:
$$
I_2^6 = -\epsilon \int_{Q} \Delta w_{x_2}|w_{x_2}|^p w_{x_2} =
\epsilon \int_{Q} \nabla w_{x_2} \cdot \nabla (|w_{x_2}|^{p-2} \, w_{x_2})
+\epsilon \int_{\Gamma} \frac{\partial w_{x_2}}{\partial n}|w_{x_2}|^{p-2}w_{x_2} d\sigma.
$$
The first term equals $\int_Q (p-1) |w_{x_2}|^{p-2} \, |\nabla w_{x_2}|^2 \,dx >0$ and the boundary term vanishes due to the boundary
condition (\ref{el_linear})$_{4,5}$.
Using the definition of $a_0(\bar w)$ (\ref{notation}) we get
\begin{equation}   \label{gammaxi}
\int_Q [ a_0(\bar w) ]_{x_i} \, w |w_{x_i}|^{p-2} w_{x_i} \leq C \, ||(\bar w+w_0)_{x_i}||_{L_p} \, ||w_{x_i}||_{L_p}^{p-1} ||w||_{W^1_p}
\leq E \, ||w||_{W^1_p}^p,
\end{equation}
thus $I_2^2 \leq E \, ||w||_{W^1_p}^p$.
In order to estimate the r.h.s we use the definition of $\tilde H$ and the H\"older inequality. We get
\begin{equation}   \label{hxi}
I_2^7 = |\int_Q \tilde H_{x_i} |w_{x_i}|^{p-2} w_{x_i} \,dx| \leq C \, ||H||_{W^1_p} ||w_{x_i}||_{L_p}^{p-1}.
\end{equation}
The important fact that we could write $H$ instead of $\tilde H$ on the r.h.s
easily results from the definition of $\tilde H$ (\ref{tildeh}).
Combining the above estimates we get
\begin{equation} \label{wx2lp}
||w_{x_2}||_{L^p}^p \leq C \big[ E \, ||\nabla w||_{L_p}^p + C \, ||H||_{W^1_p} \, ||w_{x_2}||_{L_p}^{p-1} \big].
\end{equation}
\textbf{Step 3.} In order to estimate $w_{x_1}$ we differentiate (\ref{el_linear}) with respect to $x_1$ and
multiply by $|w_{x_1}|^{p-2} \, w_{x_1}$:
$$
\underbrace{ \int_{Q} a_0(\bar w) \, |w_{x_1}|^p }_{I_3^1} + \int_Q \underbrace{ [ a_0(\bar w) ]_{x_1} |w_{x_1}|^{p-2} w_{x_1} }_{I_3^2} + \int_Q \underbrace{ w_{x_1 x_1}|w_{x_1}|^{p-2}w_{x_1} }_{I_3^3} + \underbrace{ \int_Q (\bar u+u_0) \cdot \nabla w_{x_1}|w_{x_1}|^{p-2}w_{x_1} }_{I_3^4}
$$$$
- \underbrace{ \int_{Q} \epsilon \Delta w_{x_1}|w_{x_1}|^{p-2}w_{x_1} }_{I_3^5}
= \underbrace{ \int_Q \tilde H_{x_1} \, |w_{x_1}|^{p-2}w_{x_1} - (\bar u+u_0)_{x_1} \cdot \nabla w |w_{x_1}|^{p-2}w_{x_1} }_{I_3^6}.
$$
We have
$$
I_3^3 = \frac{1}{p} \int_{Q} \partial_{x_1}|w_{x_1}|^p \,dx
= - \frac{1}{p} \int_{\Gamma_{in}} |w_{x_1}|^p \,d\sigma .
$$
Next,
$$
-I_3^5 = \epsilon \int_{Q} \nabla w_{x_1} \cdot \nabla (|w_{x_1}|^{p-2} w_{x_1}) \,dx
- \epsilon \int_{\Gamma} \frac{\partial w_{x_1}}{\partial n} |w_{x_1}|^{p-2}w_{x_1} \,d\sigma.
$$
The first term is nonnegative and the boundary term reduces to:
\begin{equation}     \label{wx1bdry1}
\epsilon \int_{\Gamma_{in}} w_{x_1 x_1} |w_{x_1}|^{p-2} w_{x_1} \,d\sigma.
\end{equation}
Note that on $\Gamma_{in}$ equation (\ref{w}) takes the form:
$$
(1+\bar u^1+u_0^1) w_{x_1} - \epsilon w_{x_1 x_1} = \tilde H|_{\Gamma_{in}}.
$$
Thus (\ref{wx1bdry1}) can be rewritten as
$$
\int_{\Gamma_{in}} [(1+\bar u^1+u_0^1)|w_{x_1}|^p - \tilde H |w_{x_1}|^{p-2}w_{x_1}] \,d\sigma.
$$
Finally,
$$
I_3^4 = -\frac{1}{p} \int_{Q} div(\bar u+u_0) |w_{x_1}|^p \,dx - \frac{1}{p} \int_{\Gamma_{in}} u_0^1 |w_{x_1}|^p \,d\sigma.
$$
Combining the above results we get
$$
C \int_{Q} |w_{x_1}|^p \,dx + \int_{\Gamma_{in}} (1-u^1-\frac{1}{p})|w_{x_1}|^p \,d\sigma \leq
$$$$
\leq \frac{1}{p} \int_{Q} div(\bar u+u_0)|w_{x_1}|^p \,dx + \int_{Q} \tilde H_{x_1} |w_{x_1}|^{p-2}w_{x_1} \,dx - \int_Q [ a_0(\bar w) ]_{x_1} w |w_{x_1}|^{p-2} w_{x_1} \,dx
$$$$
- \int_{Q} (\bar u+u_0)_{x_1} \cdot \nabla w |w_{x_1}|^{p-2}w_{x_1} \,dx + \int_{\Gamma_{in}} \tilde H |w_{x_1}|^{p-2}w_{x_1} \,d\sigma,
$$
thus using (\ref{hxi}) and (\ref{gammaxi}) we obtain
\begin{equation}  \label{wx1lp}
\begin{array}{c}
(C - E) ||w_{x_1}||_{L_p(Q)}^p
+ (1-\frac{1}{p}-E) ||w_{x_1}||_{L_p(\Gamma_{in})}^p \leq \\
\leq C\,||H||_{W^1_p} ||w_{x_1}||_{L_p(Q)}^{p-1} + E ||\nabla w||_{L_p(Q)}^p
+ ||\tilde H||_{L_p(\Gamma_{in})}||w_{x_1}||_{L_p(\Gamma_{in})}^{p-1} + E \, ||w||_{W^1_p}^p.
\end{array}
\end{equation}
\textbf{Step 4.} Combining (\ref{wx1lp}) and (\ref{wx2lp}) we get
$$
||\nabla w||_{L_p}^p + ||w_{x_1}||_{L_p(\Gamma_{in})}^p \leq
$$$$
\leq C \, [ ( ||H||_{W^1_p} + ||w||_{L_p} ) \, ||\nabla w||_{L_p}^{p-1} + ||\tilde H||_{L_p(\Gamma_{in})} ||w_{x_1}||_{L_p(\Gamma_{in})}^{p-1} ].
$$
Combining this estimate with (\ref{wlp}) we get
\begin{equation} \label{ww1p}
||w||_{L_p} + ||\nabla w||_{L_p} + ||w_{x_1}||_{L_p(\Gamma_{in})} \leq
C \, (||\tilde H||_{L_p} + ||H||_{W^1_p} + ||\tilde H||_{L_p(\Gamma_{in})}).
\end{equation}
Due to (\ref{hxi}) we have $||H||_{W^1_p}$ instead of $||\tilde H||_{W^1_p}$ on the r.h.s. and
the proof of (\ref{w-par-w1p}) is almost complete. Now it is enough to note that due to the smallness of $\bar w$ and $w_0$
in $W^1_p$ we have
$$
||\tilde H||_{L_p} \leq C ||H||_{L_p} \quad \textrm{and}
\quad ||\tilde H|_{\Gamma_{in}}||_{L_p(\Gamma|_{in})} \leq C ||H|_{\Gamma_{in}}||_{L_p(\Gamma|_{in})},
$$
thus (\ref{ww1p}) is indeed (\ref{w-par-w1p}) $\square.$

In order to complete the proof of Theorem \ref{th_w1p} we have to estimate $H$.
We will make use of the interpolation inequalities (Lemma \ref{lem_int} in the Appendix).
\begin{lem}  \label{lem_H}
Under the assumptions of Theorem \ref{th_w1p},
$\forall \delta>0$ the following estimate is valid
\begin{equation} \label{lem4_teza}
\begin{array}{c}
||H||_{W^1_p(Q)} + ||H|_{\Gamma_{in}}||_{L_p(\Gamma_{in})} \leq \\
\leq \delta ||u||_{W^2_p} + C(\delta) [||F_{\epsilon}(\bar u,\bar w)||_{L_p}+||G_{\epsilon}(\bar u,\bar w)||_{W^1_p} + ||B||_{W^{1-1/p}_p(\Gamma)} + E \, ||w||_{W^1_p}],
\end{array}
\end{equation}
where $H$ is defined in (\ref{h}).
\end{lem}
\bigskip
\textbf{Proof.} For simplicity let us denote $F:=F_{\epsilon}(\bar u, \bar w)$ and $G:=G_{\epsilon}(\bar u, \bar w)$.
Applying the interpolation inequality (\ref{int1}) to the term $||u||_{W^1_p}$ in (\ref{nablahlp}) we get:
$$
||\nabla H||_{L_p}
\leq C \big[ ||F||_{L_p} + ||G||_{W^1_p} + ||B||_{W^{1-1/p}_{p}(\Gamma)} + ||u||_{W^{1-1/p}_{p}(\Gamma)}
+\delta_1 ||u||_{W^2_p} + C(\delta_1) ||u||_{H^1} \big]
$$$$
+ E \, ||w||_{W^1_p}.
$$
In order to estimate $||H||_{L_p}$ we need to apply the interpolation inequality (\ref{int1})
and then the energy estimate (\ref{ene1}). We get
$$
||H||_{L_p}
\leq \delta_2 ||\nabla H||_{L_p}
+ C(\delta_2) \big( ||F||_{L_2} + ||G||_{L_2} + ||B||_{L_2(\Gamma)} + E \, ||w_{\epsilon}||_{W^1_p} \big).
$$
Combining the above estimates we get
\begin{equation} \label{lem4_1}
\begin{array}{c}
||\nabla H||_{L_p} + ||H||_{L_p} \leq \\
\leq (1+\delta_2)||\nabla H||_{L_p}
+ C(\delta_2) \, [||F||_{L^2} + ||G||_{L^2} + ||B||_{L^2(\Gamma)} + E \, ||w_{\epsilon}||_{W^1_p}] \leq
\\
\leq \delta_3 ||u_{\epsilon}||_{W^2_p}+C(\delta_3)
\big[ ||F||_{L_p} + ||G||_{W^1_p} + ||B||_{W^{1-1/p}_p(\Gamma)} + ||u_{\epsilon}||_{W^{1-1/p}_p(\Gamma)} + E \, ||w_{\epsilon}||_{W^1_p} \big].
\end{array}
\end{equation}
Using the trace theorem, (\ref{int1}) and (\ref{ene1}) we estimate the boundary term $||u||_{W^{1-1/p}_p(\Gamma)}$:
\begin{equation}  \label{lem4_2}
\begin{array}{c}
||u_{\epsilon}||_{W^{1-1/p}_p(\Gamma)} \leq
\delta_4 ||u_{\epsilon}||_{W^2_p} + C(\delta_4) [||F||_{L_2}+||G||_{L_2}+||B||_{L^2(\Gamma)} + E \, ||w_{\epsilon}||_{W^1_p} ].
\end{array}
\end{equation}
In order to complete the proof it is enough to estimate $H|_{\Gamma_{in}}$.
By the trace theorem we have $||div \,u||_{L_p(\Gamma_{in})} \leq C(r) ||div \,u||_{W^{1/p+r}(\Gamma_{in})} \quad \forall r>0$.
Thus, since $w|_{\Gamma_{in}}=0$, applying (\ref{int2}) and (\ref{ene1}) we get
\begin{equation}
\begin{array}{c}
||H|_{\Gamma_{in}} ||_{L_p(\Gamma_{in})} \leq ||div \, u_{\epsilon}|_{\Gamma_{in}} ||_{L_p(\Gamma_{in})} + ||G|_{\Gamma_{in}} ||_{L_p(\Gamma_{in})} \leq
\\
\leq C_{r} ||div \,u_{\epsilon}||_{W^{1/p+r}_p(Q)} + ||G||_{W^1_p(Q)}
\leq \delta_5 ||u_{\epsilon}||_{W^2_p} + C(\delta_5) ||u_{\epsilon}||_{L_p} + ||G||_{W^1_p} \leq
\\
\leq \delta_5 ||u_{\epsilon}||_{W^2_p} + C(\delta_5) [\delta_6 ||u||_{W^2_p} + C(\delta_6) ||u||_{H^1}] + ||G||_{W^1_p} \leq
\\
\leq \delta_7 ||u_{\epsilon}||_{W^2_p} + C(\delta_7) [||F||_{L_2}+||G||_{L_2}+||B||_{L^2(\Gamma)} + E \, ||w_{\epsilon}||_{W^1_p}].
\end{array}
\end{equation}
Since we control the smallness of $\delta_5$, we also control $C(\delta_5)$
and thus we can choose $\delta_6$ to make $\delta_7 = \delta_5 C(\delta_6)$ as small as we want.
Next, substituting (\ref{lem4_2}) to (\ref{lem4_1}) with (\ref{lem4_3}) we get (\ref{lem4_teza})
with $\delta$ arbitrarily small since $\delta_1 \ldots \delta_7$ can be arbitrarily small $\square.$
\\
\\
We are now ready to complete \\
\\
\emph{Proof of Theorem \ref{th_w1p}}. Let us fix $\eta>0$.
Provided that $\bar w$ and $w_0$ are small enough, combining (\ref{w-par-w1p}) and (\ref{lem4_teza}) we get
\begin{equation}    \label{th_w1p_teza}
\begin{array}{c}
||w_{\epsilon}||_{W^1_p(Q)} + ||w_{\epsilon,x_1}||_{L_p(\Gamma_{in})} \leq \\
\leq \eta ||u||_{W^2_p} +
C(\eta) [||F_{\epsilon}(\bar u, \bar w)||_{L_p} + ||G_{\epsilon}(\bar u, \bar w)||_{W^1_p} + ||B||_{W^{1-1/p}_p(\Gamma)}].
\end{array}
\end{equation}
The theory of elliptic equations applied to (\ref{el_linear})$_1$ yields
\begin{equation}
||u_{\epsilon}||_{W^2_p} \leq ||F_{\epsilon}(\bar u, \bar w)||_{L^p} + ||w_{\epsilon}||_{W^1_p}.
\end{equation}
Combining this estimate with (\ref{th_w1p_teza}) we get
$$
||u_{\epsilon}||_{W^2_p} + ||w_{\epsilon}||_{W^1_p} \leq \eta ||u_{\epsilon}||_{W^2_p} + C_\eta [ ||F_{\epsilon}(\bar u,\bar w)||_{L_p}+||G_{\epsilon}(\bar u,\bar w)||_{W^1_p}+||B||_{W^{1-1/p}_p(\Gamma)} ].
$$
Choosing for example $\eta=\frac{1}{2}$ we get (\ref{est_el_linear}). $\square$

\section{Solution of the linear system}  \label{sol_linear}
In this section we will show that the operator $T_{\epsilon}$ is well defined. We have to show that the system
(\ref{el_linear}) has a unique solution $(u,w) \in W^2_p \times W^1_p$ for
$(\bar u , \bar w) \in W^2_p \times W^1_p$ small enough. The necessary result is stated in the following
\begin{tw} \label{th_el_linear}
Assume that $||\bar u||_{W^2_p} + ||\bar w||_{W^1_p}$ is small enough. Then the system (\ref{el_linear})
has a unique solution $(u_{\epsilon},w_{\epsilon}) \in W^2_p \times W^2_p$
and the estimate (\ref{est_el_linear}) holds.
\end{tw}
We shall make here one remark concerning the above theorem.
The fact that $(u_{\epsilon},w_{\epsilon}) \in W^2_p \times W^2_p$ is a consequence
of the ellipticity of the system (\ref{el_linear}), but the estimate on $||w||_{W^2_p}$ depends
on $\epsilon$. What will be crucial for us is that (\ref{est_el_linear}) does not depend on $\epsilon$.

The system (\ref{el_linear}) has variable coefficients thus its solution is not straightforward.
In order to proove Theorem \ref{th_el_linear} we will apply the Leray-Schauder fixed-point theorem.
Given $(\bar u, \bar w) \in W^2_p \times W^1_p$ we
define an operator $S_{(\bar u, \bar w)}^{\epsilon}:W^2_p \times W^2_p \to W^2_p \times W^2_p$:\\
$(u,w) = S_{(\bar u, \bar w)}^{\epsilon}(\tilde u, \tilde w) \iff (u,w)$ is a solution to
\begin{eqnarray} \label{el_linear_2}
\begin{array}{lcr}
\partial_{x_1} u -\mu \Delta  u - (\nu + \mu) \nabla div  u + \gamma \nabla w = F^{\epsilon}_{(\bar u, \bar w)}(\tilde u, \tilde w) & \mbox{in} & Q,\\
div \,u + \partial_{x_1}w - \epsilon \Delta w = G^{\epsilon}_{(\bar u, \bar w)}(\tilde u, \tilde w) & \mbox{in}& Q,\\
n\cdot 2\mu {\bf D}( u)\cdot \tau +f \ u \cdot \tau = B
&\mbox{on} & \Gamma, \\
n\cdot  u = 0 & \mbox{on} & \Gamma,\\
w=0 & \mbox{on} & \Gamma_{in},\\
\frac{\partial w}{\partial n} = 0 & \mbox{on} & \Gamma \setminus \Gamma_{in}, \\
\end{array}
\end{eqnarray}
where
\begin{eqnarray}  \label{defFGbis}
F^{\epsilon}_{(\bar u, \bar w)}(\tilde u, \tilde w) = - (a_1(\bar w)-\gamma) \nabla \tilde w + F_{\epsilon}(\bar u,\bar w), \nonumber\\
G^{\epsilon}_{(\bar u, \bar w)}(\tilde u, \tilde w) = -(\bar w +w_0) \, div \, \tilde u - (\bar u+u_0) \cdot \nabla \tilde w + G_{\epsilon}(\bar u,\bar w).
\end{eqnarray}
We have to show that $S_{(\tilde u, \tilde w)}^{\epsilon}$ is well defined and verify that it satisfies the
assumptions of the Leray-Schauder theorem. The reason to consider $S_{(\tilde u, \tilde w)}^{\epsilon}$ on $W^2_p \times W^2_p$
instead of $W^2_p \times W^1_p$ is that it is straightforward to show its complete continuity.
\subsection{Solution of the system with constant coefficients}
In this section we show that the operator $S^{\epsilon}_{(\bar u,\bar w)}$ is well defined.
Thus we have to show that the system
\begin{eqnarray} \label{el_linear_const}
\begin{array}{lcr}
\partial_{x_1} u -\mu \Delta  u - (\nu + \mu) \nabla div  u + \gamma \nabla w = F & \mbox{in} & Q,\\
div \,u + \partial_{x_1}w - \epsilon \Delta w = G  & \mbox{in}& Q,\\
n\cdot 2\mu {\bf D}( u)\cdot \tau +f \ u \cdot \tau = B
&\mbox{on} & \Gamma, \\
n\cdot  u = 0 & \mbox{on} & \Gamma,\\
w=0 & \mbox{on} & \Gamma_{in},\\
\frac{\partial w}{\partial n} = 0 & \mbox{on} & \Gamma \setminus \Gamma_{in}, \\
\end{array}
\end{eqnarray}
where $F,G \in W^1_p(Q)$ are given functions, has a unique solution
$(u,w) \in W^2_p \times W^2_p$.  We start with showing existence
of a weak solution to the system ($\ref{el_linear_const}$). Let us recall the
definition of space $V$ (\ref{v}) and introduce another functional space
$W = \{ w \in H^1(Q): w|_{\Gamma_{in}} = 0 \}$.
Consider a bilinear form on $(V \times W)^2$:
$$
\mathbf{B}[(u,w),(v,\eta)] = \int_{Q} \{v \, \partial_{x_1} u + 2\mu \mathbf{D}(u) : \nabla v + \nu \, div \,u \, div v \} \,dx
+ \int_{\Gamma} f(u \cdot \tau) (v \cdot \tau) \,d\sigma
$$$$
 - \gamma \int_Q w \, div\,v \,dx
+ \gamma \int_Q \eta \, div \,u \,dx + \gamma \int_Q \eta \partial_{x_1} w \,dx + \gamma \epsilon \int_Q \nabla w \cdot \nabla \eta \,dx
$$
and a linear form on $(V \times W)$:
$$
\mathbf{F}(v,\eta) = \int_Q F \cdot v \,dx + \int_{\Gamma} B(v \cdot \tau) \,dx + \int_Q G \, \eta \,dx.
$$
By a weak solution to the system (\ref{el_linear_const}) we mean a couple $(u,w) \in V \times W$
satisfying
\begin{equation} \label{linear_const_weak}
\mathbf{B}[(u,w),(v,\eta)] = \mathbf{F}(v,\eta) \quad \forall (v,\eta) \in V \times W.
\end{equation}
Using the definition of $V$ and $W$ we can easily verify that
$$
\mathbf{B}[(u,w),(u,w)] \geq \int_Q 2 \mu \mathbf{D^2}(u) + \nu div^2 u \,dx + \epsilon \int_Q |\nabla w|^2 \,dx
\geq C_{\epsilon} [||u||_{H^1(Q)}+||w||_{H^1(Q)}],
$$
thus existence of the weak solution to (\ref{el_linear_const}) easily follows from the Lax-Milgram lemma.
Using standard techniques we show that the weak solution belongs to $W^2_p(Q) \times W^2_p(Q)$ and
$$
||u||_{W^2_p}+||w||_{W^2_p} \leq C_{\epsilon} \big[ ||F||_{L_p}+||G||_{W^1_p}+||B||_{W^{1-1/p}_p(\Gamma)} \big].
$$
\subsection{Complete continuity of $S^{\epsilon}_{(\bar u, \bar w)}$}
In this section we show that $S^{\epsilon}_{(\bar u, \bar w)}$ is continuous and compact. Since it is a
linear operator, it is enough to show its compactness, and this is quite obvious due to elliptic regularity
of the system $(\ref{el_linear_const})$. Namely, if we take a sequence $(\tilde u^n, \tilde w^n)$ bounded in
$W^2_p \times W^2_p$, then the sequence
$$
\big(F_{(\bar u, \bar w)}(\tilde u^n, \tilde w^n), G_{(\bar u, \bar w)}(\tilde u^n, \tilde w^n)\big)
$$
is bounded in $W^1_p \times W^1_p$. Thus the sequence
$
(u^n,w^n) = S_{(\bar u, \bar w)}^{\epsilon}(\tilde u^n, \tilde w^n)
$
is bounded in $W^3_p \times W^3_p$ (the bound on $||w||_{W^3_p}$ depends on $\epsilon$, but at this stage
$\epsilon$ is fixed, so it does not matter). The compact imbedding theorem implies that $(u^n,w^n)$ has
a subsequence that converges in $W^2_p(Q) \times W^2_p(Q)$. Thus $S_{(\bar u, \bar w)}^{\epsilon}$ is compact.

\subsection{Leray-Schauder a priori bounds}
Next we have to show a $\lambda$ - independent \textit{a priori} estimate on solutions to the equations
$(u_{\lambda},w_{\lambda}) = \lambda S^{\epsilon}_{(\bar u,\bar w)}(u_{\lambda},w_{\lambda})$, that read
\begin{eqnarray} \label{el_linear_lambda}
\begin{array}{lcr}
\partial_{x_1} u_{\lambda} -\mu \Delta  u_{\lambda} - (\nu + \mu) \nabla div  u_{\lambda} + \gamma \nabla w_{\lambda} =
\lambda F^{\epsilon}_{(\bar u, \bar w)}(u_{\lambda},w_{\lambda}) & \mbox{in} & Q,\\
div \,u_{\lambda} + \partial_{x_1}w_{\lambda} - \epsilon \Delta w_{\lambda} = \lambda G^{\epsilon}_{(\bar u, \bar w)}(u_{\lambda},w_{\lambda}) & \mbox{in}& Q,\\
n\cdot 2\mu {\bf D}(u_{\lambda})\cdot \tau +f \ u_{\lambda} \cdot \tau = B
&\mbox{on} & \Gamma, \\
n\cdot  u_{\lambda} = 0 & \mbox{on} & \Gamma,\\
w_{\lambda}=0 & \mbox{on} & \Gamma_{in},\\
\frac{\partial w_{\lambda}}{\partial n} = 0 & \mbox{on} & \Gamma \setminus \Gamma_{in}, \\
\end{array}
\end{eqnarray}
for $\lambda \in [0,1]$. Actually we should write $(u_{\lambda}^{\epsilon}, w_{\lambda}^{\epsilon})$,
but we will omit $\epsilon$ as it should not lead to any misunderstanding.
The result is stated in the following
\begin{lem}
Let $(u_{\lambda},w_{\lambda}) = \lambda S^{\epsilon}_{(\tilde u,\tilde w)}(u_{\lambda},w_{\lambda})$, then
\begin{equation}  \label{est_lambda}
||u_{\lambda}||_{W_2^p} + ||w_{\lambda}||_{W_2^p} \leq
C_\epsilon \, [ ||F(\bar u, \bar w)||_{L_p}+||G(\bar u,\bar w)||_{W^1_p}+||B||_{W^{1-1/p}_p(\Gamma)} ].
\end{equation}
\end{lem}
\textbf{Proof.} The proof is very similar to the proof theorem $\ref{th_w1p}$.
First we repeat the proof of Lemma \ref{lem_ene1} obtaining the $\lambda$-independent energy estimate
\begin{equation}  \label{ene2}
||u_{\lambda}||_{H^1}+||w_{\lambda}||_{L_2}
\leq C \big[ ||F^{\epsilon}_{(\bar u,\bar w)}(u_{\lambda},w_{\lambda})||_{L_2} + ||G^{\epsilon}_{(\bar u,\bar w)}(u_{\lambda},w_{\lambda})||_{L_2} + ||B||_{L^2(\Gamma)} \big] + E \, ||w||_{W^1_p}.
\end{equation}
Next we take the vorticity of (\ref{el_linear_lambda}):
\begin{displaymath}
\begin{array}{lcr}
\partial_{x_1} \alpha_{\lambda} - \mu \Delta \alpha_{\lambda} = rot \, ( \lambda \, F^{\epsilon}_{(\bar u,\bar v)}(u_{\lambda},w_{\lambda})) & \mbox{in} & Q, \\
\alpha_{\lambda} = - \frac{f}{\mu} (u_{\lambda} \cdot \tau) + \frac{B}{\mu} & \mbox{on} & \Gamma,
\end{array}
\end{displaymath}
where $\alpha_{\lambda} = rot \,u_{\lambda}$. Thus
\begin{displaymath}
||\alpha_{\lambda}||_{W_1^p} \leq C \left\{ ||F^{\epsilon}_{(\bar u,\bar v)}(u_{\lambda},w_{\lambda})||_{L_p(Q)} +
||B||_{W^{1-1/p}_p(\Gamma)} + ||u_{\lambda}||_{W^{1-1/p}_p(\Gamma) }
\right\}.
\end{displaymath}
Now let $u_{\lambda} = \nabla \phi_{\lambda} + A_{\lambda}^{\perp}$. Substituting this decomposition to
(\ref{el_linear_lambda}) we get
$$
-(\nu + 2\mu) \nabla \Delta \phi_{\lambda} + \nabla (\gamma w_{\lambda}) =
\lambda \, F^{\epsilon}_{(\bar u,\bar w)}(u_{\lambda},w_{\lambda}) + \mu \Delta A_{\lambda}^{\perp} - \partial_{x_{1}}A_{\lambda}^{\perp} + \partial_{x_{1}} \nabla \phi_{\lambda} -
=: \bar F_{\lambda},
$$
what can be rewritten as:
$
\nabla \left( -(\nu+2\mu)\Delta \phi_{\lambda} + \gamma w_{\lambda} \right) = \bar F_{\lambda}.
$
We denote as previously
$$
\left( -(\nu+2\mu)div \,u_{\lambda} + [ a_1(\bar w) ] w_{\lambda} \right) = \bar H_{\lambda}.
$$
Combining this identity with (\ref{el_linear_lambda})$_{2}$ we get an analog of (\ref{w}):
\begin{equation}  \label{w_lambda}
\zeta_{\lambda}(\bar w) w_{\lambda} + w_{\lambda,x_{1}} + \lambda (\bar u+u_0) \cdot \nabla w_{\lambda} - \epsilon \Delta w_{\lambda} = \tilde H_{\lambda},
\end{equation}
where
$\zeta_{\lambda}(\bar w) = \frac{\gamma}{\nu+2\mu} \, [1+\lambda(\bar w + w_0)]$ and
$\tilde H_{\lambda} = \frac{1+\lambda(\bar w+w_0)}{\nu+2\mu} \bar H_{\lambda} + \lambda G$.
Now we can repeat step by step the proof of Theorem \ref{th_w1p} obtaining the estimate
\begin{equation} \label{wlambdaw1p}
||w_{\lambda}||_{W^1_p(Q)} \leq \eta ||u_{\lambda}||_{W^2_p} + C_{\eta} [||F^{\epsilon}_{(\bar u,\bar w)}(u_{\lambda},w_{\lambda})||_{L_p}+||G^{\epsilon}_{(\bar u,\bar v)}(u_{\lambda},w_{\lambda})||_{W^1_p}+||B||_{W^{1-1/p}_p(\Gamma)}]
\end{equation}
for each $\eta>0$.
The estimates for $||u_{\lambda}||_{W^2_p}$ and $||w_{\lambda}||_{W^2_p}$
now easily result from the system (\ref{el_linear_lambda}). Namely, applying the standard elliptic theory to
(\ref{el_linear_lambda})$_1$
we obtain an estimate
\begin{equation} \label{ulambdaw2p}
||u_{\lambda}||_{W^2_p} \leq C \, \Big[ ||w_{\lambda}||_{W^1_p} + ||F^{\epsilon}_{(\bar u, \bar w)}(u_{\lambda},w_{\lambda})||_{L_p} \Big]
\end{equation}
that does not depend on $\lambda$. Next, from (\ref{el_linear_lambda})$_2$ we get an elliptic estimate
\begin{equation} \label{wlambdaw2p}
||w_{\lambda}||_{W^2_p} \leq C_{\epsilon} \, (||w_{\lambda}||_{W^1_p} + ||u_{\lambda}||_{W^2_p} + ||G^{\epsilon}_{(\bar u, \bar w)}(u_{\lambda},w_{\lambda})||_{L_p}).
\end{equation}
Combining (\ref{wlambdaw1p}), (\ref{ulambdaw2p}) and (\ref{wlambdaw2p}) we get
\begin{eqnarray} \label{ulambda_plus_wlambda_1}
||u_{\lambda}||_{W_2^p} + ||w_{\lambda}||_{W_2^p} \leq \nonumber\\
\leq C_\epsilon \, [ ||B||_{W^{1-1/p}_p(\Gamma)} + ||F^{\epsilon}_{(\bar u,\bar w)}(u_{\lambda},w_{\lambda})||_{L_p}+||G^{\epsilon}_{(\bar u,\bar w)}(u_{\lambda},w_{\lambda})||_{L_p} ],
\end{eqnarray}
but from the definition of $F^{\epsilon}_{(\bar u,\bar w)}$ and $G^{\epsilon}_{(\bar u, \bar w)}$ we have
$$
||F^{\epsilon}_{(\bar u,\bar w)}(u_{\lambda},w_{\lambda})||_{L_p}+||G^{\epsilon}_{(\bar u,\bar w)}(u_{\lambda},w_{\lambda})||_{L_p} \leq
$$$$
E \, (||u_{\lambda}||_{W^2_p} + ||w_{\lambda}||_{W^2_p}) + ||F_{\epsilon}(\bar u,\bar w)||_{L_p} + ||G_{\epsilon}(\bar u,\bar w)||_{L_p}
$$
and thus (\ref{ulambda_plus_wlambda_1}) yields (\ref{est_lambda}). $\square$

Now we are ready to complete \\
\textbf{Proof of theorem \ref{th_el_linear}}.
We have shown that the operator $S_{(\bar u, \bar w)}^{\epsilon}$ satisfies the assumptions of the Leray-Schauder theorem.
Thus there exists a fixed point $(u_{\epsilon},w_{\epsilon}) = S_{(\bar u, \bar w)}^{\epsilon}(u_{\epsilon},w_{\epsilon})$.
The fixed point is a solution to (\ref{el_linear}).
Its uniqueness follows directly from the estimate (\ref{est_el_linear}). $\square$

We have shown the existence of a unique solution to the system (\ref{el_linear}) under some smallness assumptions on
$\bar u$ and $\bar w$. Thus we define the domain ${\cal D}$ of the operator $T$:
\begin{equation}   \label{D}
{\cal D} = \big\{ (\bar u, \bar w) \in W^2_p(Q) \times W^1_p(Q): \textrm{Theorem \ref{th_el_linear} holds for} \; (\bar u ,\bar w) \big\}.
\end{equation}

\section{Solution of the regularized system}  \label{sol_reg}
In this section we show existence of a solution to an $\epsilon$-elliptic regularization to the system (\ref{system}).
The result is stated in the following
\begin{tw}  \label{th_T}
Assume that the data and $\epsilon>0$ are small enough and $f$ is large enough.
Then there exists a fixed point
$
(u^*_\epsilon, w^*_\epsilon) = T_{\epsilon}(u^*_\epsilon, w^*_\epsilon)
$
and
\begin{equation}  \label{est_main_epsilon}
||u^*_{\epsilon}||_{W^2_p} + ||w^*_{\epsilon}||_{W^1_p} \leq M,
\end{equation}
where $M$ depends on the data but does not depend on $\epsilon$ and can be arbitraily small
provided that the data is small enough.
\end{tw}
In order to prove the theorem we apply the Schauder fixed point theorem to the operator $T_{\epsilon}$
defined in (\ref{T}).
\begin{lem}
Assume that $u_0$ and $w_0$ are small enough.
Then $T_{\epsilon}(B) \subset B$ for some ball $B \subset W^2_p(Q) \times W^1_p(Q)$.
\end{lem}
\textbf{Proof}
From the definition of $F_{\epsilon}(\bar u, \bar w)$ and $G_{\epsilon}(\bar u, \bar w)$
we have
\begin{equation}  \label{FGW1p}
||F_{\epsilon}(\bar u,\bar w)||_{W^1_p} + ||G_{\epsilon}(\bar u,\bar w)||_{W^1_p} \leq
E + (||\bar u||_{W^2_p}+||\bar w||_{W^1_p})^2.
\end{equation}
Thus we can rewrite the estimate (\ref{est_el_linear}) as
\begin{equation} \label{est_el_linear_bis}
||u_{\epsilon}||_{W^2_p} + ||w_{\epsilon}||_{W^1_p} \leq C \, \big[ D + (||\bar u||_{W^2_p} + ||\bar w||_{W^1_p})^2 \big],
\end{equation}
where $D$ can be arbitrarily small provided that $||u_0||_{W^2_p}$ and $||w_0||_{W^1_p}$ are small enough.
In (\ref{est_el_linear}) we only need an estimate on $||F_{\epsilon}(\bar u,\bar w)||_{L_p}$ that hold
also for $F(\bar u,\bar w)$, but we will need the estimate in $W^1_p$ to show the compactness of $T_{\epsilon}$
and this is the reason why we introduce the regularization $F_{\epsilon}$.
Let us assume that the data is small enough to ensure $D \leq \frac{1}{4 C^2}$,
where $C$ and $D$ are the constants from (\ref{est_el_linear_bis}). Assume further that
$
||\bar u||_{W^2_p} + ||\bar w||_{W^1_p} \leq \sqrt{D}.
$
Then from (\ref{est_el_linear_bis}) we get
$$
||u_{\epsilon}||_{W^2_p}+||w_{\epsilon}||_{W^1_p} \leq 2 \, C \, D \leq \sqrt{D}.
$$
Thus $T_{\epsilon}(B) \subset B$ where $B=B(0,\sqrt{D}) \subset W^2_p(Q) \times W^1_p(Q)$ $\square.$

In the next lemma we show that $T_{\epsilon}$ is a continuous operator on $\cal D$,
where $\cal D$ is defined in (\ref{D}). The proof applies the estimate (\ref{est_el_linear})
which requires some smallness assumption,
but this assumption is also included in the definition of $\cal D$ and therefore we can
prove the continuity on the whole $\cal D$.
\begin{lem} $T_{\epsilon}$ is a continuous operator on $\cal D$.
\end{lem}
\textbf{Proof}
Let us have $(u_1,w_1) = T(\bar u_1,\bar w_1)$ and $(u_2,w_2) = T(\bar u_2,\bar w_2)$, then the
functions $u_1-u_2$ and $w_1-w_2$ satisfies the equations
\begin{eqnarray*}
\partial_{x_1} (u_1-u_2) -\mu \Delta  (u_1-u_2) - (\nu + \mu) \nabla div \, (u_1-u_2) +
\gamma (\bar w_1 +w_0 +1)^{\gamma-1} \nabla (w_1-w_2) = \\
= F_{\epsilon}(\bar u_1,\bar w_1)-F_{\epsilon}(\bar u_2,\bar w_2) - \gamma [(\bar w_1+w_0+1)^{\gamma-1} - (\bar w_2+w_0+1)^{\gamma-1}] \nabla w_2
\end{eqnarray*}
and
\begin{eqnarray*}
(\bar w_1 +w_0 +1) \, div \, (u_1-u_2) + \partial_{x_1}(w_1-w_2) + (\bar u_1+u_0) \cdot \nabla (w_1-w_2) - \epsilon \Delta (w_1-w_2) = \\
=G_{\epsilon}(\bar u_1,\bar w_1) - G_{\epsilon}(\bar u_2,\bar w_2) - (\bar w_1 - \bar w_2) \, div \,u_2 - (\bar u_1 - \bar u_2)) \cdot \nabla w_2,
\end{eqnarray*}
supplied with boundary conditions
\begin{eqnarray}
\begin{array}{lcr}
n \cdot 2\mu {\bf D}(u_1-u_2)\cdot \tau +f \, (u_1-u_2) \cdot \tau = 0
&\mbox{on} & \Gamma, \\
n\cdot  (u_1-u_2) = 0 & \mbox{on} & \Gamma,\\
w_1-w_2=0 & \mbox{on} & \Gamma_{in},\\
\frac{\partial (w_1-w_2)}{\partial n} = 0 & \mbox{on} & \Gamma \setminus \Gamma_{in}.
\end{array}
\end{eqnarray}
If $(\bar u_1, \bar w_1),(\bar u_2,\bar w_2) \in \cal D$ then
the system on $(u_1-u_2,w_1-w_2)$ satisfies the assumptions of Theorem $\ref{th_w1p}$ and thus
(\ref{est_el_linear}) yields
\begin{equation} \label{est_dif}
\begin{array}{c}
||u_1-u_2||_{W^2_p} + ||w_1-w_2||_{W^1_p} \leq \\
||F_{\epsilon}(\bar u_1,\bar w_1)-F_{\epsilon}(\bar u_2,\bar w_2)||_{L_p}
+ ||G_{\epsilon}(\bar u_1,\bar w_1)-G_{\epsilon}(\bar u_2,\bar w_2)||_{W^1_p} \\
+ ||(\bar w_1 + w_0 +1)^{\gamma-1} - (\bar w_1 + w_0 +1)^{\gamma-1} \nabla w_2 ||_{L_p}
 + ||(\bar w_1 - \bar w_2) \, div \, u_2||_{W^1_p} \\
+ ||(\bar u_1 - \bar u_2) \cdot \nabla w_2||_{W^1_p}.
\end{array}
\end{equation}
From the definition of $F_{\epsilon}(\bar u,\bar w)$ and $G_{\epsilon}(\bar u,\bar w)$ we directly get
$$
||F_{\epsilon}(\bar u_1,\bar w_1)-F_{\epsilon}(\bar u_2,\bar w_2)||_{L_p}
+ ||G_{\epsilon}(\bar u_1,\bar w_1)-G_{\epsilon}(\bar u_2,\bar w_2)||_{W^1_p}
$$$$
+ ||(\bar w_1 + w_0 +1)^{\gamma-1} - (\bar w_1 + w_0 +1)^{\gamma-1} \nabla w_2 ||_{L_p}
+ ||(\bar w_1 - \bar w_2) \, div \, u_2||_{W^1_p} \leq
$$$$
\leq C \, \big(||\bar u_1||_{W^1_p}, ||\bar w_1||_{W^1_p},||\bar u_2||_{W^1_p}, ||\bar w_2||_{W^1_p} \big) \,
\big[ ||\bar u_1- \bar u_2||_{W^2_p} + ||\bar w_1- \bar w_2||_{W^1_p} \big].
$$
In order to estimate the last term of the r.h.s. of (\ref{est_dif}) we have to use higher norm of $w_2$:
$$
||(\bar u_1 - \bar u_2) \cdot \nabla w_2||_{W^1_p} \leq C(||\bar w_2||_{W^2_p}) \, ||\bar u_1- \bar u_2||_{W^2_p}.
$$
Since on this level $\epsilon$ is fixed, we can use the elliptic regularity of the system (\ref{el_linear})
that yields
$$
||\bar w_2||_{W^2_p} \leq C_{\epsilon} ||F_{\epsilon}(\bar u_2,\bar w_2)||_{L_p} + ||G_{\epsilon}(\bar u_2,\bar w_2)||_{W^1_p} + ||B||_{L_p(\Gamma)}.
$$
Combining the above estimates we get from (\ref{est_dif}):
\begin{equation}
||u_1-u_2||_{W^2_p} + ||w_1-w_2||_{W^1_p} \leq
C_{\epsilon} \, \big[ ||\bar u_1- \bar u_2||_{W^2_p} + ||\bar w_1- \bar w_2||_{W^1_p} \big],
\end{equation}
what completes the proof of continuity of $T_{\epsilon}$ $\square.$

Now we need to proove that $T_{\epsilon}$ is a compact operator. The key is in the following lemma

\begin{lem}  \label{lemuw3pww2p}
Let us have $(u,w) = T_{\epsilon}(\bar u,\bar w)$. Then $(u,w) \in W^3_p(Q) \times W^2_p(Q)$ and
\begin{equation}  \label{uw3pww2p}
||u||_{W^3_p} + ||w||_{W^2_p} \leq C_{\epsilon} \, \big[ ||\bar u||_{W^2_p} + ||\bar w||_{W^1_p} + E \big].
\end{equation}
\end{lem}
\textbf{Proof.}
If $(u,w)=T_{\epsilon}(\bar u,\bar w)$ then in particular $w$ satisfies
$$
- \Delta w = G_{\epsilon}(\bar u, \bar w) -\partial_{x_1} w - (\bar u+u_0) \cdot \nabla w - (\bar w + w_0 +1) \, div \,u.
$$
Thus by (\ref{est_el_linear}) we have
\begin{eqnarray} \label{ww2p}
||w||_{W^2_p} \leq C_{\epsilon} \big[ ||G_{\epsilon}(\bar u,\bar w)||_{L_p} + C \, (||u||_{W^2_p} + ||w||_{W^1_p}) \big] \leq \nonumber\\
\leq C_{\epsilon} \big[ ||F_{\epsilon}(\bar u,\bar w)||_{L_p} + ||G_{\epsilon}(\bar u,\bar w)||_{W^1_p} + ||B||_{W^{1-1/p}_p(\Gamma)} \big].
\end{eqnarray}
Next, $u$ satisfies the equation
$$
-\mu \Delta u - (\nu + \mu) \nabla div u = F_{\epsilon}(\bar u, \bar w) - \partial_{x_1} u - \gamma (\bar w+w_0+1)^{\gamma-1} \nabla w,
$$
what yields
\begin{eqnarray}
||u||_{W^3_p} \leq C \big[ ||F_{\epsilon}(\bar u, \bar w)||_{W^1_p} + ||w||_{W^2_p} \big] \leq \nonumber\\
\overset{(\ref{ww2p})}{\leq} C_{\epsilon} \big[ ||F_{\epsilon}(\bar u,\bar w)||_{W^1_p} + ||G_{\epsilon}(\bar u,\bar w)||_{W^1_p} + ||B||_{W^{1-1/p}_p(\Gamma)} \big].
\end{eqnarray}
Now, from (\ref{FGW1p}) we get (\ref{uw3pww2p}). $\square$

With Lemma \ref{lemuw3pww2p} the compactness of $T_{\epsilon}$ is a straightforward consequence of the compact imbedding
theorem. Namely, if we take a sequence $(\bar u^n , \bar w^n)$ that is bounded in $W^2_p(Q) \times W^1_p(Q)$
and consider $(u^n,w^n) = T_{\epsilon}(\bar u^n, \bar w^n)$, then from (\ref{uw3pww2p}) the sequence
$(u^n,w^n)$ is bounded in $W^3_p(Q) \times W^2_p(Q)$. Thus the compact imbedding theorem implies the existence
of a subsequence $(u^{n_k}, w^{n_k})$ that converges in $W^2_p(Q) \times W^1_p(Q)$, what means that
$T_{\epsilon}$ is compact.

\textbf{Proof of theorem \ref{th_T}}. The theorem results directly from the Schauder fixed point theorem
for the operator $T_{\epsilon}$. $\square$

\section{Proof of Theorem \ref{main}} \label{sol_final}
In this section we prove our main result, Theorem \ref{main}, passing to the limit with $\epsilon$
in (\ref{el_linear}). The proof will be divided into two steps: the proof of existence of the solution
and the proof of its uniqueness. These steps are quite separated since in order to prove uniqueness
will will go back to the original system (\ref{main_system}) and modify the proof of the estimate (\ref{ene1}).

\textbf{Step 1: Existence.} Consider a decreasing sequence $\epsilon_n \to 0$. If $\epsilon_1$ is small enough
that Theorem \ref{th_T} holds (what we can assume without loss of generality), then for each
$n \in \mathbf{N}$ Theorem \ref{th_T} gives a solution $(u_{\epsilon_n},w_{\epsilon_n})$ to an
$\epsilon_n$ - elliptic regularization to (\ref{system}).

By (\ref{est_main_epsilon}) the sequence $(u_{\epsilon_n},w_{\epsilon_n})$
is uniformly bounded in $W^2_p \times W^1_p$. The compact imbedding theorem implies that
there exists a
couple $(u,w) \in W^2_p \times W^1_p$ such that (up to a subsequence)
\begin{equation} \label{limit1}
u_{\epsilon_n} \overset{W^2_p}{\rightharpoonup} u \quad \textrm{and} \quad w_{\epsilon_n} \overset{W^1_p}{\rightharpoonup} w.
\end{equation}
From the definition of ${F_\epsilon}$ and ${G_\epsilon}$ we easily get
\begin{equation} \label{limit2}
F_{\epsilon}(u_{\epsilon},w_{\epsilon}) \overset{L_p}{\to} F(u,w) \quad \textrm{and} \quad G_{\epsilon}(u_{\epsilon},w_{\epsilon}) \overset{L_p}{\to} G(u,w).
\end{equation}
%
We have to show that $(u,w)$ satisfies the system (\ref{system}). Clearly we have
\begin{equation} \label{limit3}
\begin{array}{lr}
\Delta u_{\epsilon_n} \overset{L_p}{\rightharpoonup} \Delta u, &
\nabla div \, u_{\epsilon_n} \overset{L_p}{\rightharpoonup} \nabla div \,u , \\
\partial_{x_1} w_{\epsilon_n} \overset{L_p}{\rightharpoonup} \partial_{x_1} w , &
\nabla w \overset{L_p}{\rightharpoonup} w.
\end{array}
\end{equation}
Thus it remains to show convergence in nonlinear terms, but this is also straightforward. We have $\forall \phi \in L_q:$
$$
\int_{Q} \phi (w_{\epsilon} + w_0 +1)^{\gamma-1} \nabla w_{\epsilon} \,dx =
$$$$
\int_{Q} \phi [(w_{\epsilon} + w_0 +1)^{\gamma-1} - (w + w_0 +1)^{\gamma-1}] \nabla w_{\epsilon}
+ \int_Q \phi (w + w_0 +1)^{\gamma-1} \nabla w_{\epsilon} \,dx
$$
Since $\phi (w + w_0 +1)^{\gamma-1} \in L_q$,
the second term converges to $\int_Q \phi (w + w_0 +1)^{\gamma-1} \nabla w \,dx$.
The first term
$$
\Big|\int_{Q} \phi [ (w_{\epsilon} + w_0 +1)^{\gamma-1} - (w + w_0 +1)^{\gamma-1} ] \, \nabla w_{\epsilon} \,dx \Big| \leq
$$$$
\leq ||\phi [(w_{\epsilon} + w_0 +1)^{\gamma-1} - (w + w_0 +1)^{\gamma-1}]||_{L_q} ||w_{\epsilon}||_{W^1_p}
\overset{\epsilon \to 0}{\to} 0,
$$
since by the compact imbedding theorem $w_{\epsilon} \overset{L_q}{\to} w \quad \forall \; 1 \leq q < +\infty$.
Thus
\begin{equation} \label{limit4}
\int_{Q} \phi (w_{\epsilon} + w_0 +1)^{\gamma-1} \nabla w_{\epsilon} \,dx \to \int_Q \phi (w + w_0 +1)^{\gamma-1} \nabla w \,dx.
\end{equation}
Similarily we can show that
\begin{equation} \label{limit5}
(w_{\epsilon} + w_0 +1) \, div \,u_{\epsilon} + (u_{\epsilon}+u_0) \cdot \nabla w_{\epsilon}
\overset{L_p}{\rightharpoonup} (w + w_0 +1) \, div \,u + (u+u_0) \cdot \nabla w .
\end{equation}
From (\ref{limit2}), (\ref{limit3}), (\ref{limit4}) and (\ref{limit5}) we see that $(u,w)$ satisfies
(\ref{system})$_{1,2}$ a.e. in $Q$. The trace theorem implies that
\begin{equation}
\begin{array}{lcr}
w_{\epsilon}|_{\gamma_{in}} \overset{L_p(\Gamma_{in})}{\rightharpoonup} w|_{\Gamma_{in}}, &
u|_{\Gamma} \overset{L_p(\Gamma)}{\to} u|_{\Gamma}, &
\mathbf{D}(u) \overset{L_p(\Gamma_{in})}{\rightharpoonup} u_{\Gamma}.
\end{array}
\end{equation}
Thus $u$ satisfies (\ref{system})$_{3,4}$ a.e. on $\Gamma$ and $w$ satisfies (\ref{system})$_5$ a.e. on $\Gamma_{in}$.
Now take $v = u + u_0 + \bar v$ and $\rho = w + w_0 + \bar \rho$, where $u_0$ and $w_0$ are
extensions to the boundary data defined in (\ref{extension}) and $(\bar v , \bar \rho) \equiv ([1,0],1)$
is the constant solution.
Then $(v,\rho)$ satisfies the system (\ref{main_system}).

In order to show the estimate (\ref{est_main}) we repeat the proof of Theorem \ref{th_w1p}
obtaining
\begin{equation} \label{est_main1}
||u||_{W^2_p} + ||w||_{W^1_p} \leq C \, \big[ ||F(u,w)||_{L_p} + ||G(u,w)||_{W^1_p} + ||B||_{W^{1-1/p}_p(\Gamma)} \big].
\end{equation}
We have
\begin{equation} \label{est_main2}
||F(u,w)||_{L_p} + ||G(u,w)||_{W^1_p} \leq D + \big(||u||_{W^2_p} + ||w||_{W^1_p} \big)^2,
\end{equation}
where $D$ can be arbitrarily small provided that the data is small enough.
From (\ref{est_main1}) and (\ref{est_main2}) we conclude (\ref{est_main}).

\textbf{Step 2: Uniqueness.} In order to prove the uniqueness of the solution
in a class of small perturbations to of the constant flow $(\bar v, \bar \rho)$
consider
$(v_1,\rho_1)$ and $(v_2,\rho_2)$ both being solutions to (\ref{main_system})
satisfying the estimate (\ref{est_main}).
We will apply the ideas of the proof of the energy estimate (\ref{ene1})
in order to show that
\begin{equation} \label{est_dif0}
||v_1 - v_2||_{H^1}^2 + ||\rho_1 - \rho_2||_{L_2}^2 = 0.
\end{equation}
For simplicity let us denote the differences $u:=v_1-v_2$ and $w:=\rho_1-\rho_2$.
We will follow the notation of constants introduced before, namely $E$ shall denote a constant
dependent on the data that can be arbitrarily small provided that the data is small enough, whereas
$C$ will denote a constant dependent on the data that is controlled, but not necessarily small.
In order to show (\ref{est_dif0}) it is enough to prove that
\begin{equation} \label{est_dif1}
||u||_{H^1} \leq E ||w||_{L_2}
\end{equation}
and
\begin{equation} \label{est_dif2}
||w||_{L_2} \leq C ||u||_{H^1}.
\end{equation}
If we substract the equations on $(v_1,\rho_1)$ and $(v_2,\rho_2)$ there appears a term
$\rho_1^{\gamma} - \rho_2^{\gamma}$.
We will use the fact that
$\rho_1,\rho_2 \sim 1 \Rightarrow \rho_1^{\gamma} - \rho_2^{\gamma} \sim \gamma (\rho_1 - \rho_2)$,
more precisely, we can write:
$$
\rho_1^{\gamma} - \rho_2^{\gamma} = (\rho_1 - \rho_2) \underbrace{ \int_0^1 \gamma [t \rho_1 + (1-t) \rho_2]^{\gamma-1} \,dt }_{I_\gamma}
$$
and we have $I_{\gamma} \simeq \gamma$. Now we easily verify that the difference $(u,w)$ satisfies the system
\begin{equation}  \label{dif}
\begin{array}{c}
w \, v_2 \cdot \nabla v_2 + \rho_1 \, u \cdot \nabla v_2 + \rho_1 \, v_1 \cdot \nabla u
- \mu \Delta u - (\mu+\nu) \nabla div \, u + I_\gamma \nabla w = 0,  \\
\rho_1 \, div \, u + w \, div \, v_2 + u \cdot \nabla \rho_2 + v_1 \cdot \nabla w = 0, \\
n \cdot 2 \mu \mathbf{D}(u) \cdot \tau|_{\Gamma} = 0, \\
n \cdot u|_{\Gamma} = 0, \\
w|_{\Gamma_{in}} = 0.
\end{array}
\end{equation}
We modify the proof of (\ref{ene1}),
multiplying (\ref{dif})$_1$ by $\rho_1 \, u$ and integrate over $Q$
(the reason why we take $\rho_1 \, u$ instead of $u$ will be explained soon). We get
$$
\int_Q (2 \mu \mathbf{D^2}(u) + \nu \rho_1 \, div^2 \, u)\,dx
+ \underbrace{ \int_Q \big[ (\rho_1-1) \mathbf{D}(u) : \nabla u + \mathbf{D}(u) : (u \otimes \nabla \rho_1) \big] \,dx }_{I_1}
$$$$
- I_{\gamma} \int_Q w \, \rho_1 \, div \, u \,dx
+ \int_{\Gamma} \rho_1 \, f \, u^2 \, d\sigma
$$$$
- \underbrace{ \int_Q w \, u \, \nabla \rho_1 }_{I_2}
+ \underbrace{ \int_Q \rho_1^2 u^2 \cdot \nabla v_2 \,dx }_{I_3}
+ \underbrace{ \int_Q u \, w \, \rho_1 \, v_2 \cdot \nabla v_2 \,dx }_{I_4}
+ \underbrace{ \int_Q  \rho_1^2 \, (v_1 \cdot \nabla u) \cdot \ u \,dx }_{I_5} = 0.
$$
We have
$
|I_1|+|I_2|+|I_3|+|I_4| \leq E \, (||u||_{H^1}^2 + ||w||_{L_2}^2).
$
Now let us split $I_5$ into two parts:
$$
2 I_5 = \underbrace{ \int_Q (\rho_1^2 \, v_1^{(1)} - 1) \,
\partial_{x_1} |u|^2 + \rho_1^2 \, v_1^{(2)} \, \partial_{x_2} |u|^2 \,dx}_{I_5^1}
+ \underbrace{ \int_Q \partial_{x_1} |u|^2 \,dx}_{I_5^2}.
$$
We have
$|I_5^1| \leq E ||u||_{H^1}^2$
and
$
I_5^2 = \int_{\Gamma} |u|^2 n^{(1)} \, d\sigma.
$
The last term can be integrated by parts and combined with the boundary term involving friction.
Thus applying the Korn inequality (\ref{Korn}) we get
$$
C \, ||u||_{H_1}^2 + \int_{\Gamma} (\rho_1 \, f + n^{(1)}) |u|^2 \, d\sigma - I_{\gamma} \int_Q w \, div u \,dx \leq E \, ||u||_{H^1}^2.
$$
For the friction coefficient $f$ large enough the boundary term will be positive and thus
\begin{equation}  \label{est_dif3}
||u||_{H^1}^2 \leq C \int_Q w \, \rho_1 \, div\, u \,dx.
\end{equation}
The reason why we multiplied (\ref{dif})$_1$ by $\rho_1 \, u$ is that now we have this function on the
r.h.s of (\ref{est_dif0}) instead of $div \, u$.
In order to derive (\ref{est_dif1}) from (\ref{est_dif3}) we express $\rho_1 \, div \,u$ in terms of $w$
using the equation (\ref{dif})$_2$. Thus we can rewrite (\ref{est_dif3}) as
\begin{equation}
||u||_{H^1}^2 \leq - \underbrace{ \int_Q w^2 \, div \, v_2 \,dx }_{I_6}
- \underbrace{ \int_Q w \, v_1 \cdot \nabla w \,dx }_{I_7}
- \underbrace{ \int_Q w \, u \cdot \nabla \rho_2 \,dx }_{I_8}.
\end{equation}
Obviously $|I_6| \leq E ||w||_{L_2}^2$
and, since $p>2$, we have $|I_8| \leq ||\nabla \rho_2||_{L_p} \, ||w||_{L_2} \, ||u||_{L_q}$
for some $q < \infty$. Thus from the imbedding theorem we get
$
|I_8| \leq E (||w||_{L_2}^2 + ||u||_{H^1}^{2}).
$
Integrating by parts in $I_7$ and using the boundary conditions we get
$$
- 2 I_5 = \int_Q w^2 \, div \, v_1 \,dx - \int_{\Gamma_{out}} v_1^{(1)} \,d\sigma.
$$
The boundary term is positive since $v_1^{(1)} \sim 1$, thus
$
- I_5 \leq C \, ||\nabla v_1||_{\infty} ||w||_{L_2}^2 = E ||w||_{L_2^2}.
$
Combining the estimates for $I_4$, $I_5$ and $I_6$ we get (\ref{est_dif1}).

Now in order to complete the proof we have to show (\ref{est_dif2}).
Note that it is useless to multiply (\ref{dif})$_2$ by $w$ since we would obtain a term
$w^2 div \, v_2$. Thus we adapt again the approach from
the proof of (\ref{ene1}) and write an expression on a pointwise value of $w^2$:
$$
w^2(x_1,x_2) = \int_0^{x_1} w \, w_s (s,x_2) ds =
- \int_0^{x_1} \frac{\rho_1}{v_1^{(1)}} w \, div \,u \,dx
$$$$
- \int_0^{x_1} \frac{1}{v_1^{(1)}}  \Big( w^2 \, div \,v_2 + w \, u \cdot \nabla \rho_2 \Big) \,dx
- \int_0^{x_1} \frac{v_1^{(2)}}{v_1^{(1)}} w \, \partial_{x_2} w \,dx =: w_1^2 + w_2^2 + w_3^2.
$$
Note that we have $\rho_1,v_1^{(1)} \sim 1$ and thus $\forall \delta>0$:
\begin{equation}  \label{w_31}
\int_Q w_1^2 \,dx \leq C \, (||w||_{L_2} \, ||div \,u||_{L_2}) \leq \delta \, ||w||_{L_2}^2 + C(\delta) \, ||u||_{H^1}^2.
\end{equation}
Next we easily get
$
\int_Q w_3^2 \, dx \leq E (||w||_{L_2}^2 + ||u||_{H^1}^2),
$
and we only have to deal with $w_3^2$. We have
$
\int_Q w_3^2 \,dx = \int_0^1 \big[ \int_{P_{x_1}} w_3^2 \,dx \big] \,dx_1.
$
Consider the inner integral
$$
\int_{P_{x_1}} w_3^2 \,dx =
- \int_{P_{x_1}} \partial_{x_2} \frac{v_1^{(2)}}{v_1^{(1)}} w^2 \,dx
+ \int_{\partial P_{x_1}} w^2 v_1^{(1)} v_1^{(2)} n^{(2)} \,d\sigma .
$$
The boundary term vanishes and thus
\begin{equation}
\int_Q w_3^2 \leq C ||\partial_{x_2} \frac{v_1^{(2)}}{v_2^{(1)}}||_{\infty} ||w||_{L_2}^2 \leq E ||w||_{L_2}^2.
\end{equation}
Choosing for example $\delta=\frac{1}{2}$ in (\ref{w_31}) we get (\ref{est_dif2}),
what completes the proof of (\ref{est_dif0}). We have shown that the solution is unique, and thus
comleted the proof of theorem \ref{main}. $\square$

\section{Appendix}
\begin{lem} (Korn inequality) Let $V = \{v \in H^1(Q): (n \cdot v)|_{\Gamma} = 0 \}$. Then
$\exists C = C(Q)$:
\begin{equation} \label{Korn}
\int_{Q} 2 \mu {\bf D^2}(u) + \nu div^2 u \,dx \geq C_Q \| u \|_{W_2^1}^2.
\end{equation}
\end{lem}
The proof can be found in (\cite{PM1}, Lemma 2.1) or in (\cite{Sol}, Lemma 4).
\begin{lem} (Helmoltz decomposition)
There exists a couple of functions $(\phi,A) \in (W^2_p)^2$ such that
$n \cdot \nabla^{\perp} A|_{\Gamma} =0$
\begin{equation}
u_{\epsilon}= \nabla \phi + \nabla^{\perp} A. \label{Helm}
\end{equation}
Moreover,
\begin{equation}  \label{est_helm}
||\phi||_{W^2_p}  + ||A||_{W^2_p} \leq C \, ||u||_{W^1_p}.
\end{equation}
\end{lem}
The proof can be found in \cite{Ga}. The last auxiliary result we need are
the interpolation inequalities.
\begin{lem} (interpolation inequalities): \label{lem_int} \\
$\forall \epsilon >0 \quad \exists C(\epsilon,p,Q)$ such that $\forall f \in W^1_p(Q)$:
\begin{equation}  \label{int1}
||f||_{L_p} \leq \epsilon ||\nabla f||_{L_p} + C(\epsilon,p,Q) ||f||_{L_2}.
\end{equation}
and $\forall \, \eta>0$ such that $\frac{1}{p}+\eta<1$, $\forall \epsilon >0 \quad \exists C(\eta,\epsilon,p,Q)$
such that $\forall f \in W^1_p(Q)$:
\begin{equation}   \label{int2}
||f||_{W^{1/p+\eta}_p} \leq \epsilon ||f||_{W^1_p} + C(\epsilon,p,Q) ||f||_{L_p}.
\end{equation}

\end{lem}
\textbf{Proof.} Inequality (\ref{int1}) results from the following inequality (\cite{Ad}, Theorem 5.8):
\begin{equation}  \label{int1_1}
||f||_{L_p} \leq K \, ||f||_{W^1_2}^{\theta} \, ||f||_{L_2}^{1-\theta}
\end{equation}
for each $2 \leq p < \infty$, where $\theta = \frac{n(p-2)}{2p}$ and $K=K(p,Q)$.
Using Cauchy inequality with $\epsilon$ we get \ref{int1}.

The inequality (\ref{int2}) for the functions defined on $\mathbf{R^n}$ is a well-known result
from the theory of Besov spaces (\cite{Tr}). It can be extended for the square domain due to its symmetry.
$\square.$

\smallskip

\textbf{Acknowledgements.} The author would like to thank Piotr Mucha and
Milan Pokorny for fruitful discussions and remarks concerning
this paper.


\begin{thebibliography}{99}
\footnotesize
\bibitem{Ad} R.Adams, J.Fournier, \emph{Sobolev spaces}\/, 2nd ed., Elsevier, Amsterdam, 2003
\bibitem{Ga} G.P.Galdi, \emph{An Introduction to the mathematical theory of the Navier-Stokes Equations}\/,
				Vol.I, Springer-Verlag, New York, 1994
\bibitem{GaNoPa} G.P.Galdi, A.Novotny, M.Padula,
			\emph{On the two-dimensional steady-state problem of a viscous gas in an exterior domain}\/,
			Pacific J. Math. 179,1 (1997), 65-100.
\bibitem{Fu} H.Fujita, \emph{Remarks on the Stokes flow under slip and leak boundary conditions of friction type}\/,
			Topics in Mathematical Fluid Mechanics, 73-94, Quad.Mater. 10 (2002)
\bibitem{Gi} D.Gilbarg, N.S.Trudinger, \emph{Elliptic Partial Differential Equations of Second Order}\/,
				2nd ed., Springer-Verlag, Berlin, 1983
\bibitem{GM} O.Glass, P.B.Mucha, \emph{Inviscid limit for the 2-D stationary Euler system with arbitrary force
				in simply connected domains}\/, Applicationes Mathematicae 35,1 (2008), 49-67
\bibitem{Kw1} R.B.Kellogg, J.R.Kweon, \emph{Compressible Navier-Stokes equations in a bounded domain
				with inflow boundary condition}\/, SIAM J.Math.Anal. 28,1(1997), 94-108
\bibitem{Kw2} R.B.Kellogg, J.R.Kweon, \emph{Smooth Solution of the Compressible Navier-Stokes Equations
				in an Unbounded Domain with Inflow Boundary Condition}\/,
				J.Math.Anal. and App. 220 (1998), 657-675
\bibitem{Kw3} J.R.Kweon, M.Song, \emph{Boundary geometry and regularity of solution to the compressible
				Navier-Stokes equations in bounded domains of $\mathbf{R^n}$}\/,
				ZAMM Z.Angew.Math.Mech. 86,6 (2006), 495-504
\bibitem{PM1} P.B.Mucha, \emph{On Navier-Stokes equations with Slip Boundary Conditions
			in an Infinite Pipe}\/, Acta Applicandae Mathematicae 76(2003), 1-15
\bibitem{PMMP} P.B.Mucha, M.Pokorny, \emph{On a new approach to the issue of existence and regularity
			for the steady compressible Navier-Stokes equations}\/, Nonlinearity 19(2006), 1747-1768
\bibitem{PM3} P.B.Mucha, R.B.Rautmann, \emph{Convergence of Rothe's scheme for the Navier-Stokes equations
			with slip boundary conditions in 2D domains}\/, ZAMM Z.Angew.Math.Mech., 86,9(2006), 691-701
\bibitem{No} A.Novotny, \emph{Some remarks to the compactness of steady compressible isentropic Navier-Stokes equations
			via the decomposition method}\/, Comment.Math.Univ.Carolinae 37,2(1996), 305-342
\bibitem{NoS} A.Novotny, I.Straskraba, \emph{Mathematical Theory of Compressible Flows}\/,
			Oxford Science Publications, Oxford 2004
\bibitem{PS1} P.I.Plotnikov, J.Sokolowski, \emph{Domain dependence of solutions to Compressible Navier-Stokes
			Equations}\/, SIAM J.Control Optim. 45,4, 1165-1197
\bibitem{PS2} P.I.Plotnikov, J.Sokolowski, \emph{On Compactness, Domain Depedence and Existence of
			Steady State Solutions to Compressible Isothermal Navier-Stokes equations},
			J.Math.Fluid.Mech. 7(2005), 529-573
\bibitem{PS3} P.I.Plotnikov, J.Sokolowski, \emph{Stationary Solutions of Navier-Stokes equations
				for diatomic gases}\/, Russian Math Surveys 62:3, 561-593
\bibitem{Sol} V.A.Solonnikov, V.E.Scadilov, \emph{On a boundary value problem for a stationary system
				of Navier-Stokes equations}, Trudy Mat.Inst.Steklov. 125(1973), 186-199
\bibitem{Te} R.Temam, \emph{Navier Stokes Equations}\/, North-Holland, Amsterdam, 1977.
\bibitem{Tr} H.Triebel, \emph{Interpolation theory, function spaces, differential operators.}\/,
			North-Holland Mathematical Library, 18. North-Holland Publishing Co., Amsterdam-New York, 1978
\bibitem{VZ} A.Valli, W.M.Zajaczkowski, \emph{Navier - Stokes equations for compressible fluids: global existence and qualitative properties of the solutions in the general case}\/,
 			Comm. Math. Phys. 103,2 (1986), 259-296
\end{thebibliography}
\end{document}